\documentclass[journal,10pt,twoside]{IEEEtran} 

\usepackage{amsthm}
\usepackage{amsfonts}
\usepackage{amssymb}
\usepackage{mathrsfs}
\usepackage{mathtools}
\usepackage{float}
\usepackage[pdftex]{graphicx}
\usepackage{amsmath}
\usepackage{color}
\usepackage{multirow,multicol}
\usepackage{graphicx}
\usepackage{times}
\usepackage{textcomp}
\usepackage{verbatim}
\usepackage[table]{xcolor}
\usepackage{balance}
\usepackage{lipsum}
\usepackage[inline]{enumitem}
\usepackage{cuted}
\usepackage[caption=false,font=footnotesize]{subfig}
\usepackage{cite}
\usepackage{algpseudocode,algorithm}
\usepackage{hyperref}
\usepackage{tcolorbox}
\usepackage{setspace,epstopdf}
\usepackage[table]{xcolor}
\definecolor{color1}{RGB}{199,209,232}
\definecolor{color2}{RGB}{230,231,233}

\DeclareMathOperator*{\maximize}{maximize\;} 
\DeclareMathOperator*{\minimize}{minimize\;} 
\DeclareMathOperator*{\subjectto}{subject\;to\hspace{3pt}} 

\makeatletter
\newcommand{\smallsym}[2]{#1{\mathpalette\make@small@sym{#2}}}
\newcommand{\make@small@sym}[2]{%
	\vcenter{\hbox{$\m@th\downgrade@style#1#2$}}%
}
\newcommand{\downgrade@style}[1]{%
	\ifx#1\displaystyle\scriptstyle\else
	\ifx#1\textstyle\scriptstyle\else
	\scriptscriptstyle
	\fi\fi
}
\makeatother

\begin{document}
	
	\title{ Twenty-Five Years of Advances in Beamforming:
		From Convex and Nonconvex Optimization to Learning Techniques}
	\author{\IEEEauthorblockN{ \normalsize
			Ahmet M. Elbir, \textit{Senior Member, IEEE}, Kumar Vijay Mishra, \textit{Senior Member, IEEE}, \\ Sergiy A. Vorobyov, \textit{Fellow, IEEE}, and Robert W. Heath, Jr., \textit{Fellow, IEEE} }
		\thanks{A. M. Elbir is with the Interdisciplinary Centre for Security, Reliability and Trust, University of Luxembourg, Luxembourg; and  Duzce University, Duzce, Turkey  (e-mail: ahmetmelbir@ieee.org).}
		\thanks{K. V. Mishra is with the United States DEVCOM Army Research Laboratory, Adelphi, MD 20783 USA (e-mail: kvm@ieee.org).}
		\thanks{S. A. Vorobyov is with the Department of Signal Processing and Acoustics, Aalto University, Espoo 02150 Finland (e-mail: sergiy.vorobyov@aalto.fi).}
		\thanks{R. W. Heath Jr. is with the Department of Electrical and Computer Engineering, North Carolina State University, Raleigh, NC 27606 USA (e-mail: rwheathjr@ncsu.edu).}
	}

	\maketitle
		\IEEEpeerreviewmaketitle

	\begin{abstract}
		Beamforming is a signal processing technique to steer, shape, and focus an electromagnetic wave using an array of sensors toward a desired direction. It has been used in several engineering applications such as radar, sonar, acoustics, astronomy, seismology, medical imaging, and communications. With the advent of multi-antenna technologies in, say, radar and communication, there has been a great interest in designing beamformers by exploiting convex or nonconvex optimization methods. Recently, machine learning is also  leveraged for obtaining attractive solutions to more complex beamforming scenarios. This article captures the evolution of beamforming in the last twenty-five years from convex-to-nonconvex optimization and optimization-to-learning approaches. It provides a glimpse into these important signal processing algorithms for a variety of transmit-receive architectures, propagation zones, propagation paths, and multi-disciplinary applications.
		
	\end{abstract}

	
	\section{Introduction}
	
	Beamforming is ubiquitous and essential to a multitude of array processing applications such as radar, sonar, acoustics, astronomy, seismology, ultrasound, and communications \cite{beamformingVanVeen1988Apr}. Recent advances in mobile communications, usage of large arrays, high-frequency sensors, near-field signal recovery, and smart radio environments open up interesting and novel signal processing problems in beamforming. These applications are driving the need for higher robustness, flexible deployment, and low complexity in beamforming algorithms and an emphasis on advanced signal processing that should be tailored for emerging application-specific requirements.

	Early experiments with beamforming could be traced back to Guglielmo Marconi, who used a circular array with four antennas to improve the gain of trans-Atlantic Morse code transmission in 1901~\cite{marconi_simons1996guglielmo}. A similar early demonstration of gains provided by a phased array to direct radio waves was in 1905 by Karl Ferdinand Braun, who shared Nobel Prize in physics with Marconi in 1909 for their contributions to wireless telegraphy \cite{sarkar2006history}. In the 1940s, antenna diversity as a technique to overcome fading was developed for phased array radars and radio astronomy \cite{bartlett1941dual}. By the 1950-1960s, with the development of phased arrays for sonars, the steering of signals with antenna arrays was no longer restricted to electromagnetic waves \cite{chen2016beamforming}.
	
	\textit{Adaptive beamforming}~\cite{Capon1969Aug,widrow1967Dec} emerged in the late 1960s, wherein a processor at the antenna back-end updates and compensates the array weights. In particular, Bernard Widrow introduced the least-mean-square (LMS) algorithm to update the weights at every iteration by estimating the gradient of the mean squared error (MSE) between the desired and received signals~\cite{widrow1967Dec}. Subsequently, J. Capon proposed selecting the weight vectors or \textit{beamformers} to minimize the array output power. The \textit{Capon beamformer} is subjected to the linear constraint that the signal-of-interest (SoI) does not suffer from any distortion, e.g., direction mismatch, signal fading, local scattering, etc.~\cite{Capon1969Aug,VorobyovBookChapter_2014Jan}. Hence, this technique is also usually referred to as the minimum variance (MV) distortionless response (MVDR) beamforming.

	The performance of Capon beamformer strongly depends on the knowledge of SoI, which is imprecise in practice because of the differences between the assumed and true array responses. The beamforming performance is usually measured by the signal-to-interference-plus-noise ratio (SINR). This may severely degrade even in the presence of small errors or mismatches in the steering vector~\cite{VorobyovBookChapter_2014Jan}.   
	In the past, numerous approaches were proposed  to improve the robustness against errors/mismatches in the look direction~\cite{pointingError_Cox2005,generalizedSidelobe_Jablon1986Aug}, array manifold~\cite{mismodeling_Gershman1995Oct}, and local scattering~\cite{localScattering_Astely1999Dec}.
	These techniques were limited to only the specific mismatch they treat~\cite{worstCase_Vorobyov2003Jan}, thereby giving rise to early generalization of
	\textit{robust beamforming} approaches such as sample matrix inversion (SMI) algorithm~\cite{robustAdaptiveBF_Cox1987}, robust Capon beamforming~\cite{robustCapon_Li2003Jun}, eigenspace-based beamformer~\cite{eigenspacedBF_Feldman1994Apr}, worst-case performance optimization~\cite{worstCase_Vorobyov2003Jan} and general-rank beamformer~\cite{generalRank_Shahbazpanahi2003Aug,generalRank2_Khabbazibasmenj2013Sep}. 
	
	In the late 1990s and early 2000s, significant progress was made toward robust beamformer design by exploiting \textit{convex optimization}~\cite{convexBF_Gershman2010Apr}. These methods typically consider minimizing the effect of mismatches in the array steering vectors  and the look direction based on the worst-case performance optimization~\cite{worstCase_Vorobyov2003Jan, minimumVariance_Lorenz2005Apr,robustCapon_Li2003Jun}. Here, the optimization problem is cast as a second-order cone (SOC) program and efficiently solved by interior-point methods. It may also be desirable to design a robust MVDR beamformer by including the uncertainty in the array manifold via an ellipsoid or a sphere model for a particular look direction~\cite{robustCapon_Li2003Jun,minimumVariance_Lorenz2005Apr}. 
	
	During the late 2000s, certain applications of beamforming that have nonconvex objective functions or constraints gained salience. These included robust adaptive beamforming with additional constraints related to the positive semi-definiteness (PSD) of the signal covariance matrix~\cite{generalRank2_Khabbazibasmenj2013Sep}, norm of the steering vectors~\cite{doublyConstrained_Li2004Aug,HuangMe19,HassMe08,ArashHassMe12}, and stochastic distortionless response~\cite{ProbMe1,ProbHuangMe22}; multicast transmit beamforming~\cite{multicaseBF_Sidiropoulos2006Jun}, and hybrid (analog/digital) beamforming~\cite{hybridBF_Heath_Ayach2014Jan}. The solution to these \textit{nonconvex optimization} problems usually requires recasting the problem into a 
	tractable form through the use of, for example, semi-definite relaxation (SDR), compressed sensing (CS)~\cite{hybridBF_Heath_Ayach2014Jan}, and alternating optimization~\cite{convexBF_Gershman2010Apr}. \textcolor{black}{Solving for beamforming weights is generally considered as a \textit{continuous optimization} problem. However, there is a smaller body of literature \cite{Savas2022Nov,multiCasting_Alternating_Demir2014Sep} on \textit{discrete/combinatorial} techniques. Here, the beamforming weights are selected from a set of exponentials with discretized angles. }

	\begin{figure*}[t]
		\centering
		{\includegraphics[draft=false,width=\textwidth]{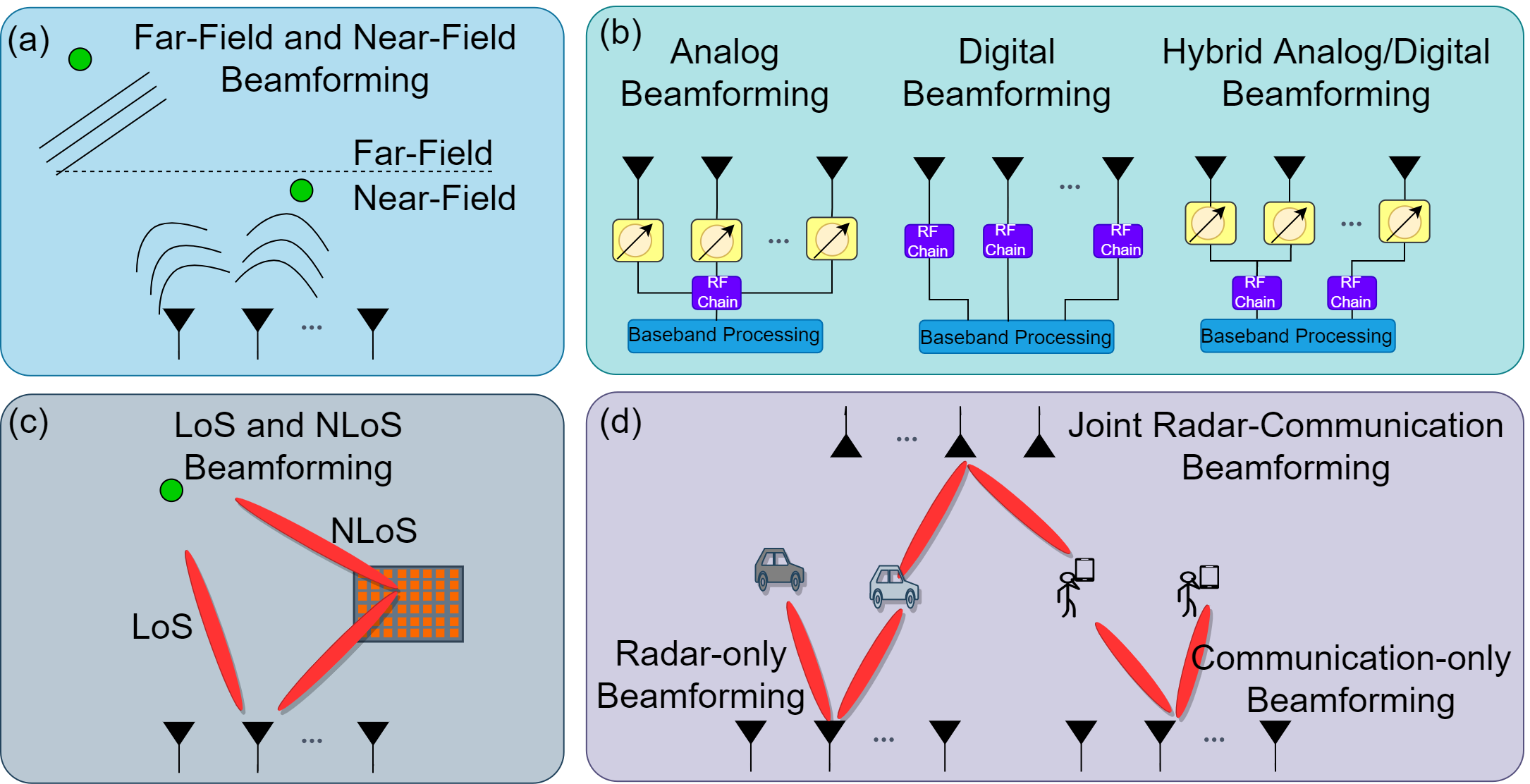} } 
		\caption{Major classes of beamforming methods by (a) transmission range:  far- and near-fields; (b) transceiver architectures: analog, digital, and hybrid beamforming; (c) paths: LoS and NLoS beamforming, wherein the NLoS path is controlled via joint active (transmitter) and passive (intelligent reflecting surface) devices; (d) applications: radar, communications, and joint radar-communications.
		}
		\label{fig_Beamforming}
	\end{figure*}

	In the last decade, with the advent of new cellular communications technologies, beamforming has been extensively investigated for multi-antenna systems~\cite{ hybridBF_Heath_Ayach2014Jan}. The fourth generation (4G) networks (2009-present) operating at $2.2$-$4.9$~GHz, use up to $32$ antennas in a multiple-input multiple-output (MIMO) configuration. 5G systems (2019-present) offer support for larger antenna arrays as well as communication at frequencies above $24$ GHz. Support for larger arrays is essential in millimeter wave systems to overcome shrinking antenna sizes~\cite{heath_Overview_hybridBF}. To reduce the hardware, cost, power, and area in mmWave massive MIMO systems, \textit{hybrid} (analog and digital) beamforming has been introduced~\cite{hybridBF_Heath_Ayach2014Jan, heath_Overview_hybridBF}. Unlike a conventional digital beamformer employing a single RF chain dedicated to each antenna, hybrid approaches employ a few (large) radio-frequency (RF) chains (analog components, e.g., phase shifters) to reduce the hardware cost. The hybrid beamformer design is also nonconvex because of unit-modulus constraint owing to the use of phase shifters in the analog beamformers. This problem has been addressed through techniques such as sparse matrix reconstruction via CS~\cite{hybridBF_Heath_Ayach2014Jan}, optimization over Riemannian manifolds~\cite{hybridBFAltMin}, phase-extraction~\cite{sohrabi_JSTSP}, and Gram-Schmidt orthogonalization~\cite{alkhateeb2016frequencySelective}.
	
	Very recently, data-driven methods such as \textit{machine learning} (ML) have been leveraged to obtain beamformers. ML is a subset of artificial intelligence (AI) that allows neural networks (NNs) to learn directly from precedents, data, and examples without being explicitly programmed. \textcolor{black}{Many beamformers involve nonlinear operations. In this context, NNs are particularly attractive because they successfully approximate non-linear functions or predict the class of a function that is divided by a non-linear decision boundary.} Compared to the model-based techniques, ML has lower post-training computational complexity, expedited design procedure, and robustness against imperfections/mismatches~\cite{deepL_robustBeamforming_Mohammadzadeh2022Oct,elbir2021DecAFamily, elbirQuantizedCNN2019}.  The ML-based hybrid beamforming is also envisioned as a key to realize massive MIMO architectures beyond 5G communications~\cite{hb_DL_GYLi}, such as 6G systems operating at Terahertz (THz) bands. \textcolor{black}{This is largely because ML is helpful in processing copious amounts of antenna array data generated by massive MIMO systems employed at higher frequencies.    }
	
	To shed light on the evolution of beamforming techniques, this article  presents an overview of the aforementioned approaches while focusing on major breakthroughs during the last 25 years. Specifically, the article aims at: i) highlighting the two significant leaps in this research, i.e., convex-to-nonconvex optimization, and optimization-to-learning-based beamforming; ii) depicting in detail the analytical background and the relevance of signal processing tools for beamforming, and iii) introducing the major challenges and emerging signal processing applications of  beamforming. Fig.~\ref{fig_Beamforming} summarizes some important classes of beamformers discussed in this article.

	\textit{Notation:} Throughout this paper, uppercase and lowercase bold letters denote matrices and vectors, respectively. Also, $(\cdot)^\textsf{T}$ and $(\cdot)^{\textsf{H}}$ denote the transpose and conjugate transpose operations, respectively. For a matrix $\mathbf{A}\in \mathbb{C}^{M\times N}$ and a vector $\mathbf{a}\in \mathbb{C}^{N}$; $[\mathbf{A}]_{ij}$  $[\mathbf{A}]_k$, $\Re\{\mathbf{A}\}$ and $\Im\{\mathbf{A} \}$, and $a_i$ correspond to the $(i,j)$-th entry, $k$-th column, the real and imaginary parts of $\mathbf{A}$, and the $i$-th entry of $\mathbf{a}$, while $\mathbf{A}^{\dagger}$ denotes the Moore-Penrose pseudo-inverse of $\mathbf{A}$, and $\mathbf{I}$ is the identity matrix of proper size. \textcolor{black}{ $\|\mathbf{a} \|_2 = (\sum_{i = 1}^{N} |a_i |^2)^{\frac{1}{2}}  $  and $\| \mathbf{A}\|_\mathcal{F} = (\sum_{i = 1}^{M} \sum_{j=1}^{N} |[\mathbf{A}]_{ij}|^2 )^{\frac{1}{2}} $ denote the $l_2$-norm and Frobenius norm, respectively.     }

	\section{Convex Optimization for Beamforming}
	\label{sec:ConvexOpt}
	Convex optimization recasts originally difficult-to-design beamformers to computationally attractive problems that yield exact or approximate solutions through algorithms such as interior-point methods. Its applications have traditionally transcended from simple exact Capon approach to more complex transmit, multicast, network, and distributed beamformers; see, e.g., \cite{convexBF_Gershman2010Apr} and references therein for details. In the following, we summarize the techniques that yield exact solutions. The approximate solutions are considered under nonconvex beamformers in the sequel.
	\subsection{Capon beamformer}
	Consider an antenna array with $N$ elements. Define $\mathbf{a}(\theta) \in\mathbb{C}^{N}$ as the array response to a plane-wave narrowband SoI $s(t_i)$, $i = 1,\cdots,T$, where $T$ is the number of snapshots, arriving from the direction-of-arrival (DoA) angle $\theta$. In particular, the steering vector $\mathbf{a}(\theta) $ is  
	\begin{align}
	\mathbf{a}(\theta) = \frac{1}{\sqrt{N}}[1, e^{-\mathrm{j}2\pi \frac{d}{\lambda}\sin \theta }, \cdots, e^{-\mathrm{j}2\pi \frac{(N-1)d}{\lambda}\sin \theta}]^\textsf{T},
	\end{align}
	where $d$ is the element spacing and $\lambda$ is the wavelength. Then, the $N\times 1$ antenna array output is 
	\begin{align}
	\mathbf{y}(t_i)= \mathbf{a}(\theta) s(t_i) + \mathbf{e}(t_i),
	\end{align}
	where $\mathbf{e}(t_i)\in \mathbb{C}^{N}$ denotes the temporarily and spatially white Gaussian noise vector with variance $\sigma^2$.	
	
	The received signals are multiplied by the beamforming weights i.e., $w_1,\cdots, w_N \in \mathbb{C}$. Therefore, the combined beamformer output becomes
	\begin{align}
	y_o(t_i) = \mathbf{w}^\textsf{H} \mathbf{y}(t_i) = \mathbf{w}^\textsf{H}\mathbf{a}(\theta)s(t_i) + \mathbf{w}^\textsf{H}\mathbf{e}(t_i),
	\end{align}
	where $\mathbf{w} = [w_1,\cdots,w_N]^\textsf{T}$ includes the beamformer weights. To recover the signal $s(t_i)$, the beamformer weights are optimized via
	\begin{align}
	\label{optCapon1}
	\minimize_{\mathbf{w}} \mathbf{w}^\textsf{H} \mathbf{R}_{y} \mathbf{w} \hspace{20pt} \subjectto \mathbf{w}^\textsf{H}\mathbf{a}(\theta) = 1,
	\end{align}
	where $\mathbf{R}_{y} = \frac{1}{T} \sum_{i = 1}^{T}\mathbf{y}(t_i)\mathbf{y}^\textsf{H}(t_i)$ is the sample covariance matrix of the array output. The optimal solution for (\ref{optCapon1}) yields the Capon beamformer~\cite{Capon1969Aug}:
	\begin{align}
	\label{optCaponSolution}
	\mathbf{w}_\mathrm{opt} = \left(\mathbf{a}^\textsf{H}(\theta)\mathbf{R}_{y}^{-1}\mathbf{a}(\theta) \right)^{-1}  \mathbf{R}_{y}^{-1} \mathbf{a}(\theta).
	\end{align}
	\textcolor{black}{This beamformer requires the knowledge of $\mathbf{a}(\theta)$ and  $\mathbf{R}_y$. Therefore, its performance depends on the accuracy of the steering vector constructed from the estimate of $\theta$ as well as the sample covariance matrix $\mathbf{R}_y$. }
	
	To stabilize the mainbeam response in the presence of pointing error~\cite{pointingError_Cox2005}, additional \textcolor{black}{constraints} are added to the optimization problem as
	\begin{align}
	\label{generalizedSidelobeCanceler}
	\minimize_{\mathbf{w}} \mathbf{w}^\textsf{H} \mathbf{R}_{y} \mathbf{w} \hspace{20pt} \subjectto \mathbf{C}^\textsf{H}\mathbf{w} = \mathbf{u},
	\end{align}
	where $L$ many constraints are represented by $\mathbf{C}\in \mathbb{C}^{L\times N}$ and $\mathbf{u}\in \mathbb{C}^{L}$. For example, if it is desired to maximize the beampattern at $30^\circ$ and place a null at $40^\circ $, then $\mathbf{C} = [\mathbf{a}(30^\circ), \mathbf{a}(40^\circ)]^\textsf{T}$ and $\mathbf{u} = [1, 0]^\textsf{T}$. The solution to this constrained problem is $\mathbf{w}_\mathrm{C} = \mathbf{R}_y^{-1} \mathbf{C} (\mathbf{C}^\textsf{H}\mathbf{R}_y^{-1}\mathbf{C})^{-1}\mathbf{u}$~\cite{generalizedSidelobe_Jablon1986Aug}.

	\subsection{Loaded SMI beamformer}
	Even in the ideal case, wherein the SoI direction $\theta$ is accurately known, beamforming performance may significantly deteriorate because of a small training sample size $T$. This is mitigated by adding a regularization term $\gamma$ to the objective function in (\ref{optCapon1}) leading to \textit{loaded SMI (LSMI) beamforming}~\cite{robustAdaptiveBF_Cox1987}:
	\begin{align}
	\label{optLoadedSMI}
	\minimize_{\mathbf{w}} \mathbf{w}^\textsf{H} \mathbf{R}_{y} \mathbf{w} + \gamma ||\mathbf{w} ||_2 \hspace{20pt} \subjectto \mathbf{w}^\textsf{H}\mathbf{a}(\theta) = 1.
	\end{align}
	Its solution is $\mathbf{w}_\mathrm{LSMI} = \mathbf{R}_\mathrm{LSMI}^{-1}\mathbf{a}(\theta)$, where $ \mathbf{R}_\mathrm{LSMI} =  \mathbf{R}_{y} + \gamma\mathbf{I}_N$.

	\subsection{Robust Capon beamformer}
	The exact knowledge of the SoI direction $\theta$ required by Capon beamformer is not available in practice. This is addressed by \textit{robust beamforming}, which provides tolerance against the inaccuracies in estimated SoI direction and the corresponding steering vector. A \textit{robust} variant of Capon beamforming was introduced in~\cite{robustCapon_Li2003Jun}, wherein the convex optimization problem is 
	\begin{align}
	\label{robustCaponBF}
	\minimize_{\mathbf{w}} \mathbf{w}^\textsf{H} \mathbf{R}_{y}^{-1}\mathbf{w},\hspace{20pt} \subjectto \|  \mathbf{w} - \bar{\mathbf{a}} \|_2\leq \epsilon,
	\end{align}
	where $\bar{\mathbf{a}} = \mathbf{\mathbf{a}}(\theta + \Delta_\theta)$ is the inaccurate steering vector for mismatched direction $\theta + \Delta_\theta$. 
	
	\subsection{Beamforming with worst-case performance optimization}
	A more general approach is considered in~\cite{worstCase_Vorobyov2003Jan} by taking into account the distortions in the steering vector as $\tilde{\mathbf{a}} = \mathbf{a}(\theta) + \Delta_\mathbf{a}$, where $\Delta_\mathbf{a}\in \mathbb{C}^N$ represents the steering vector distortions. As a result, the optimization problem is based on the worst-case beamforming performance.  Relying on the bounded Euclidean norm as $||\Delta_\mathbf{a}||_2 \leq \varepsilon$ corresponding to the case of spherical uncertainty~\cite{worstCase_Vorobyov2003Jan}, the following convex problem is formulated:
	\begin{align}
	\label{optSpherical}
	\minimize_{\mathbf{w}} \mathbf{w}^\textsf{H} \mathbf{R}_{y}\mathbf{w},\hspace{20pt} \subjectto  |\mathbf{w}^\textsf{H} \tilde{\mathbf{a}}| \geq  1,  \|\Delta_\mathbf{a} \|_2 \leq \varepsilon,
	\end{align}
	for which the LSMI-based solutions may also be obtained~\cite{VorobyovBookChapter_2014Jan,convexBF_Gershman2010Apr}. A similar approach, called \textit{robust MV beamforming}, introduced in~\cite{minimumVariance_Lorenz2005Apr}, is based on ellipsoidal uncertainty. \textcolor{black}{Both spherical (e.g., $\|\tilde{\mathbf{a}} - \mathbf{a}(\theta) \|_2 \leq \varepsilon$ in (9)) and ellipsoidal (e.g., $(\tilde{\mathbf{a}} - \mathbf{a})^\textsf{H} \mathbf{V} (\tilde{\mathbf{a}} - \mathbf{a}) \leq \tilde{\varepsilon}$, where $\mathbf{V}\in \mathbb{C}^{N\times N}$ is a PSD matrix) models are used to ensure robust solutions. The latter may naturally lead to a more accurate uncertainty description~\cite{minimumVariance_Lorenz2005Apr} than that with spherical models~\cite{Beck2007Jan,minimumVariance_Lorenz2005Apr}, if more information than just the same uncertainty radius in all mismatch dimensions is available and an uncertainty ball is replaced by an uncertainty ellipsoid. Assuming the availability of more information about the mismatch is, however, somewhat contradictory to the notion of robustness.   }
	
	{\color{black}The structure of the beamformer design problem also depends on the noise model. Some beamforming techniques are based on the MV criterion mentioned earlier. However, this criterion is statistically optimal only when the SoI, interference, and noise are Gaussian. The non-Gaussian case leads to a nonconvex problem as
		\begin{align}
		\label{optNonGaussian}
		\minimize_{\mathbf{w}} ||\mathbf{Y}^\textsf{H}\mathbf{w} ||_p^p   , \hspace{20pt} \subjectto \mathbf{a}^\textsf{H}(\theta)\mathbf{w} = 1,
		\end{align}
		where $\mathbf{Y} = [\mathbf{y}(t_1),\cdots, \mathbf{y}(t_T)]\in \mathbb{C}^{N\times T} $, and $\|\mathbf{y}(t_i)\|_p^p = \left(\sum_{n=1}^{N} y_n(t_i) \right)^{1/p} $ denotes the $\ell_p$-norm for $p \geq 1$. Note that (\ref{optNonGaussian}) reduces to Capon beamforming of (4) for $p=2$. The solution for (\ref{optNonGaussian}) is achieved via iterative reweighted MVDR techniques~\cite{minimumDispersion_Jiang2014Feb}. In addition to generalizing the noise model, a specific choice of priors over the distribution of the beamforming weights may also be used in, say, sparsity-driven beamforming \cite{Parayil2021May}.
	}
	
	\subsection{Beamforming for general-rank source}
	In practice, the source signal is incoherently scattered such that the point-source assumption may not hold~\cite{generalRank_Shahbazpanahi2003Aug} and the array covariance matrix is no longer $\mathrm{rank}$-$1$. Therefore, instead of a constraint on a single steering vector, the SoI covariance matrix is used. The corresponding MVDR-type optimization problem is
	\begin{align}
	\label{generalRank}
	\minimize_{\mathbf{w}} \mathbf{w}^\textsf{H}\mathbf{R}_{y}\mathbf{w} \hspace{20pt} \subjectto \mathbf{w}^\textsf{H}\mathbf{R}_s\mathbf{w} = 1, 
	\end{align}
	where  $\mathbf{R}_s$ is  SoI covariance matrix~\cite{generalRank2_Khabbazibasmenj2013Sep}. The optimal solution to (\ref{generalRank}) is	
	$\mathbf{w}_\mathrm{GR} = \mathcal{P}\left[\mathbf{R}_{y}^{-1}\mathbf{R}_s \right]$, where $\mathcal{P}\left[\cdot \right]$ is the principal eigenvector operator.

	\section{Nonconvex Beamformer Design}
	\label{sec:NonConvexOpt}
	Nonconvex beamformers \cite{doublyConstrained_Li2004Aug, multicaseBF_Sidiropoulos2006Jun,oneBit_CE_Li2017May, hybridBF_Heath_Ayach2014Jan, HassMe08, ArashHassMe12, HuangMe19, ProbMe1, ProbHuangMe22} tackle the design problem by recasting or relaxing it into tractable convex forms. This may be achieved by dropping the nonconvex constraints, or decoupling the beamforming design into multiple convex subproblems.
	
	\subsection{PSD-constrained beamforming}
	The general-rank beamforming solution in (\ref{generalRank}) requires the knowledge of signal covariance matrix $\mathbf{R}_{ s}$, which is not always available~\cite{generalRank_Shahbazpanahi2003Aug,generalRank2_Khabbazibasmenj2013Sep}. The actual signal correlation matrix is then not guaranteed to be PSD and usually modeled as $\tilde{\mathbf{R}}_{ s} = \mathbf{R}_{\rm s} + \boldsymbol{\Delta}_{ s}$. \textcolor{black}{To guarantee PSD-ness of $\tilde{\mathbf{R}}_{ s}$, decompose it as $\tilde{\mathbf{R}}_{ s} = \mathbf{QQ}^\textsf{H}$ with the mismatch parameter  $\boldsymbol{\Delta}_Q$ bounded as $||\boldsymbol{\Delta}_{ Q}||_2 \leq \varepsilon_{ Q}$. The resulting nonconvex problem is 
		\begin{align}
		&\minimize_{\mathbf{w}} \max_{||\boldsymbol{\Delta}_{ y} ||_2 \leq \varepsilon_{ y}} \mathbf{w}^\textsf{H} (\mathbf{R}_{ y} + \boldsymbol{\Delta}_{ y} ) \mathbf{w}  \nonumber\\
		&
		\quad\subjectto \min_{||\boldsymbol{\Delta}_{ Q}||_2 \leq \varepsilon_{ Q}} \mathbf{w}^\textsf{H} (\mathbf{Q} + \boldsymbol{\Delta}_{ Q})^\textsf{H} (\mathbf{Q} + \boldsymbol{\Delta}_{ Q}) \mathbf{w} \geq 1, \label{psdConstrained}
		\end{align}
		where  $\boldsymbol{\Delta}_{ y}$, with $||\boldsymbol{\Delta}_{ y}||_2 \le \varepsilon_{ y}$, represents the mismatch in  $\mathbf{R}_{ y}$.} The efficient solution to the nonconvex problem in (\ref{psdConstrained}) is obtained via polynomial-time difference-of-convex functions (POTDC) algorithm~\cite{generalRank2_Khabbazibasmenj2013Sep}. 
	
	\subsection{Norm-constrained beamforming based on steering vector estimation}
	Apart from the uncertainty constraint (\ref{robustCaponBF}) of robust Capon beamformer \cite{robustCapon_Li2003Jun}, \cite{doublyConstrained_Li2004Aug}  considers an additional norm constraint for beamformer weights in a more general setting as
	\begin{align}
	\label{normConstrainedBF}
	\minimize_{\mathbf{w}} \mathbf{w}^\textsf{H} \hat{\mathbf{R}}^{-1}\mathbf{w} \quad \subjectto  \| \mathbf{a} - \tilde{\mathbf{a}} \|_2 \leq \epsilon_a, \| \mathbf{a}\|_2^2 = N, 
	\end{align}
	which is identical to (\ref{robustCaponBF}) and convex without the constraint $\| \mathbf{a}\|_2^2 = N$. The nonconvex problem in (\ref{normConstrainedBF}) is called \textit{doubly-constrained} robust Capon beamforming~\cite{doublyConstrained_Li2004Aug}. It is iteratively solved by interpreting the optimization as a covariance fitting problem. Thus, a robust  beamformer is obtained by robustly estimating the array steering vector. This formulation was further improved in \cite{HassMe08}, where the difference between the actual and presumed steering vectors is iteratively estimated without making any assumption  on either the norm of the mismatch vector or its probability distribution. 
	
	The solution developed in \cite{HassMe08} has led to a formulation in \cite{ArashHassMe12} of a new constraint, which guarantees that an estimate of the source steering vector does not converge to any steering vectors of interference signals as well as their linear combinations. This steering vector estimation problem is 
	\begin{align} \label{Converted}
	\minimize_{ \hat{\mathbf a}  } \hat{\mathbf a}^\textsf{H} \hat{\mathbf R}^{-1} \hat{\mathbf a} \quad \subjectto \left\| \hat{\mathbf a} \right\|^2_2 = N, \; 
	\hat{\mathbf a}^\textsf{H}  \tilde{\mathbf C} \hat{\mathbf a} \leq \Delta_0,
	\end{align}
	where the last constraint is new; $\hat{\mathbf a}\in\mathbb{C}^N$ is the estimate of ${\mathbf a}$; $\tilde{\mathbf C} = \int_{\tilde{\Theta}} {\mathbf a (\theta ) \mathbf a^\textsf{H} (\theta ) \, d \theta}$ and $\tilde{\Theta}$ is the complement of the angular sector $\Theta =[\theta_{\min}, \theta_{\max}]$ where the desired signal is located; and $\Delta_0$ is a uniquely selected value for a given $\Theta$, that is, $\Delta_0 \triangleq \max_{\theta \in \Theta} \mathbf a^\textsf{H} (\theta ) \tilde{\mathbf C} \mathbf a (\theta )$, representing the boundary line to distinguish approximately whether or not the direction of $\mathbf a$ is in the actual signal angular sector $\Theta$.
	
	To account for gain perturbations in the steering vector, \cite{HuangMe19} added double-sided norm constraint to the problem \eqref{Converted} as
	\begin{align}
	\label{optRegion}
	&\minimize_{ \hat{\mathbf a} } \hat{\mathbf a}^\textsf{H} \hat{\mathbf R}^{-1} \hat{\mathbf a} 
	\nonumber\\
	&\quad \subjectto \; \hat{\mathbf a}^\textsf{H} {\mathbf C} \hat{\mathbf a} \geq
	\Delta_1, \nonumber\\
	&\quad N (1 - \eta_1)\le\| \hat{\mathbf a} \|^2_2 \leq N (1 + \eta_2), \nonumber\\
	&
	\quad\|{\mathbf V}^\textsf{H} (\hat{\mathbf a} - {\mathbf a}_0)\|^2_2 \leq \epsilon_u,
	\end{align}
	where ${\mathbf a}_0 = {\mathbf a} (\theta_0)$, $\theta_0 = (\theta_{\max} + \theta_{\min})/2$ is the middle value of the region $\Theta$; ${\mathbf V}\in\mathbb{C}^{N \times N}$ denotes a generalized similarity constraint together with ${\mathbf a}_0$ and $\epsilon_u$; ${\mathbf C} = \int_{\Theta} {\mathbf a (\theta ) \mathbf a^\textsf{H} (\theta ) \, d \theta}$; and $\Delta_1$, $\eta_1$, and $\eta_2$ are selected values. In (\ref{optRegion}), the generalized similarity condition implies that imperfect knowledge of the desired steering vector $\hat{\mathbf a}$ is described as in a convex set (in particular, an ellipsoidal set when ${\mathbf V}$ is of full row rank).
	
	All these problems are nonconvex, but can be often exactly solved through SDR, iterative SOC program, quadratic matrix inequality (QMI), and bilinear matrix inequality (BLMI) approaches.
	
	\subsection{Chance-constrained beamforming}
	In many applications, it is more natural that the distortionless constraint is satisfied with a certain probability. This leads to the chance-constrained robust adaptive beamforming problem \cite{ProbMe1}:
	\begin{align}
	\minimize_{{\mathbf w}}\, {\mathbf w}^\textsf{H} \hat{\mathbf R} {\mathbf w} \quad \subjectto \; {\rm Pr} \{ |
	{\mathbf w}^\textsf{H} \tilde {\mathbf a} | \geq 1 \} \geq p,
	\label{MVDRprob}
	\end{align}
	where $p$ is a certain pre-selected probability value, and ${\rm Pr} \{ \cdot \}$ stands for the probability operator. This problem corresponds to minimizing the beamformer output power subject to the {\it stochastic} constraint that the probability of the signal distortionless response is greater than or equal to some selected value $p$. The constraint may also be viewed as a {\it non-outage probability constraint} where the outage probability $p_{\rm out}=1-p$ is defined as that of violating the inequality $|{\mathbf w}^\textsf{H} \tilde {\mathbf a} | \geq 1$ for random $\tilde {\mathbf a}$ that consists of a presumptive steering vector and the mismatch that is assumed to be random. Problem \eqref{MVDRprob} is nonconvex and specified by the mismatch distribution. The solutions of \eqref{MVDRprob} for the case of Gaussian distributed mismatch of the signal steering vector and for the worst-case distribution are well approximated by the corresponding SOC programs \cite{ProbMe1}. 
	
	In \cite{ProbHuangMe22}, chance-constrained nonconvex formulation of robust adaptive beamforming considers a more practical scenario, wherein both interference-plus-noise covariance (INC) matrix ${\mathbf R}_{i+n}$ and the true steering vector $\mathbf a$ are not precisely known. \textcolor{black}{It also shows chance-constrained beamformer to have higher output SINR than other convex (LSMI) and nonconvex (worst-case optimization) beamformers~\cite{ProbHuangMe22}}. Considering both ${\mathbf R}_{i+n}$ and $\mathbf a$ as random variables, the robust adaptive beamforming becomes
	\begin{align} \label{rob-beamf-DRO}
	&\minimize_{{\mathbf w}}\, \max_{G_1 \in{\cal S}_1} E_{G_1} \{{\mathbf w}^\textsf{H} {\mathbf R}_{i+n} {\mathbf w} \} 
	\nonumber\\
	& \subjectto \; \min_{G_2\in{\cal S}_2} E_{G_2} \{{\mathbf w}^\textsf{H} {\mathbf a} {\mathbf a}^\textsf{H} {\mathbf w}\} \geq 1,
	\end{align}
	where $E_{G_1}\{\cdot\}$ $\left( E_{G_2}\{\cdot\} \right)$ denotes the statistical expectation under the distribution $G_1$ ($G_2$) and ${\cal S}_1$ (${\cal S}_2$) is a set of distributions $G_1$ ($G_2$) for random matrix ${\mathbf R}_{i+n}$ (random vector $\mathbf a$) as, respectively,
	\begin{equation} \label{distribution-set-D1}
	{\cal S}_1=\left\{G_1 \in {\cal M}_1~\left|~\begin{array}{l} {\rm Pr}_{G_1} \{{\mathbf R}_{i+n} \in {\cal Z}_1\} = 1\\ E_{G_1} \{{\mathbf R}_{i+n}\} \succeq {\mathbf 0} \\ \|E_{G_1} \{{\mathbf R}_{i+n}\} - {\mathbf S}_0\|_\mathcal{F} \leq \rho_1 \end{array} \right. \right\},
	\end{equation}
	and
	\begin{equation}
	{\cal S}_2=\left\{G_2 \in{\cal M}_2~\left|~\begin{array}{l} {\Pr}_{G_2}\{{\mathbf a} \in {\cal Z}_2 \} = 1 \\ E_{G_2}\{\mathbf{a}\} = {\mathbf a}_0 \\ E_{G_2}\{{\mathbf a} {\mathbf a}^\textsf{H}\} = {\boldsymbol \Sigma} + {\mathbf a}_0 {\mathbf a}_0^\textsf{H} \end{array}\right.\right\}, 
	\end{equation}
	where ${\cal M}_1$ and ${\cal M}_2$ are sets of all probability measures, ${\cal Z}_1$ and ${\cal Z}_2$ are Borel sets, ${\mathbf S}_0$ is the empirical mean of ${\mathbf R}_{i+n}$, that is, the sample covariance matrix ${\mathbf R}_y$, and ${\Pr}_{G_1}\{\cdot\}$ is the probability of an event under the distribution $G_1$. Assume the mean ${\mathbf a}_0$ and covariance matrix ${\boldsymbol \Sigma} \succ {\mathbf 0}$ of random vector ${\mathbf a}$     under the true distribution $\bar G_2$ are known. Then, the set ${\cal S}_2$ includes all probability distributions on ${\cal Z}_2$ which have the same first- and second-order moments as $\bar G_2$. This problem is called {\it distributionally robust beamforming} because it considers distributional uncertainty in both the steering vector and ${\mathbf R}_{i+n}$.

	\subsection{Multicast transmit beamforming}
	In wireless communications, multicast beamforming is used for broadcasting data streams $s(t_i)$ toward multiple radio receivers. Consider a transmitter with $N$-element antenna array that aims to deliver a signal to $U$ single-antenna users. Denote the wireless channel between the transmitter and the $u$-th receiver by $\mathbf{h}_u\in\mathbb{C}^{N} $. Then, for the beamformed transmitted signal $\mathbf{x}(t_i) = \mathbf{w}s(t_i)$,  the received signal at the $u$-th user is $	{y}_u(t_i) = \mathbf{h}_u^\textsf{H}\mathbf{x}(t_i) + e_u(t_i)$,	where $e_u(t_i)$ is the noise signal with variance $\sigma_u^2$. Then, the multicast beamforming problem is \cite{multicaseBF_Sidiropoulos2006Jun}
	\begin{align}
	\label{multiCastProblem1}
	&\minimize_{\mathbf{w}} \| \mathbf{w}\|_2 \nonumber\\
	& |\mathbf{w}^\textsf{H}\tilde{\mathbf{h}}_u | \geq 1, \hspace{20pt} u\in \{1,\cdots, U\},
	\end{align}
	where $\tilde{\mathbf{h}}_u = \mathbf{h}/\sqrt{\rho_{\mathrm{min},u} \sigma_u^2 }$ is the normalized channel vector with the minimum received SNR $\rho_{\mathrm{min},u}$ and the noise variance $\sigma_u^2$ for the $u$-th receiver. The  optimization in (\ref{multiCastProblem1}) is a quadratically constrained quadratic programming (QCQP) problem with nonconvex constraints. A rigorous solution is based on reformulating the problem using SDR. To this end, define an $N\times N$ rank-1 matrix $\mathbf{M} = \mathbf{ww}^\textsf{H}$. Then, the rank constraint is removed to recast the problem in a convex form  as
	\begin{align}
	&\minimize_{\mathbf{M}}  \mathrm{trace}\{\mathbf{M} \}
	\nonumber\\
	&\quad
	\subjectto \mathrm{trace}\{\mathbf{M}\mathbf{D}_u  \} \geq  1, \hspace{20pt} \mathbf{M} \succeq \mathbf{0},
	\end{align}
	where $\mathbf{D}_u = \tilde{\mathbf{h}}_u \tilde{\mathbf{h}}_u^\textsf{H}$ and the beamformer weight is obtained via eigenvalue decomposition of $\mathbf{M}$. A more accurate solution to (\ref{multiCastProblem1}) is obtained by rewriting $\mathbf{M} = \mathbf{w}_1 \mathbf{w}_2^\textsf{H}$ and then alternatingly solving for $\mathbf{w}_1$ and $\mathbf{w}_2$ using an iterative procedure until convergence~\cite{multiCasting_Alternating_Demir2014Sep}.

	\begin{figure*}[t]
		\centering
		{\includegraphics[draft=false,width=\textwidth]{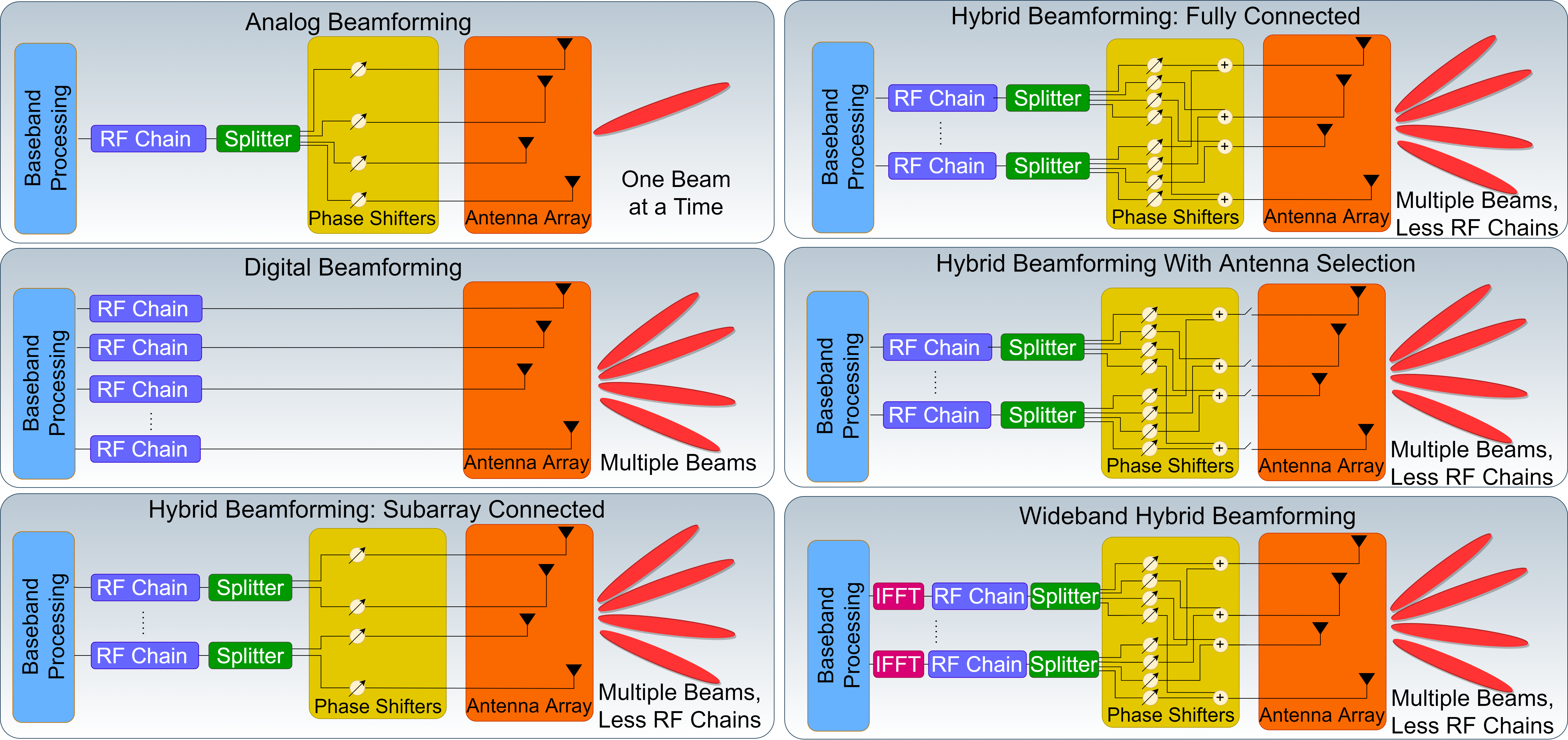} } 
		\caption{Transmitter architectures for analog (Top Left), digital \textcolor{black}{(Middle Left)}, and hybrid beamforming. Analog beamforming generates only one beam because it employs a single RF chain. On the other hand, multiple beams are obtained via digital beamformers, but at the cost of multiple RF chains. It is possible to generate multiple beams with fewer RF chains in the \textit{hybrid} approach through configurations such as subarray-connected (Bottom Left), fully-connected (Top Right), sparse antenna-selective (Middle Right) and wideband (Bottom Right) architectures. 
		}
		\label{fig_AnalogDigitalBeamforming}
	\end{figure*}

	\vspace{-12pt}
	\subsection{Hybrid analog/digital beamforming}
	Compared to analog- and digital-only-beamformers, hybrid analog/digital beamforming architecture may have a lower hardware cost while also providing satisfactory spectral efficiency (SE) and multiple beams (Fig.~\ref{fig_AnalogDigitalBeamforming}). In fact, for massive antenna array processing applications such as 5G communications, hybrid beamforming has emerged as the preferred means to realize large arrays with only a moderate increase in baseband signal processing~\cite{heath_Overview_hybridBF,sohrabi_JSTSP}.
	
	Consider a hybrid beamforming scenario, wherein the transmitter employs $N$ antennas and $N_\mathrm{RF}$ RF chains to send $N_\mathrm{S}$ data streams. Denote the analog and digital beamformers by matrices  $\mathbf{F}_\mathrm{RF} \in \mathbb{C}^{N\times N_\mathrm{RF}}$ and $\mathbf{F}_\mathrm{BB} \in \mathbb{C}^{N_\mathrm{RF}\times N_\mathrm{S}}$, respectively. Here, each element of $\mathbf{F}_\mathrm{RF}$ has constant-modulus because they are realized by phase-shifters, i.e., $[\mathbf{F}_\mathrm{RF}]_{i,j}={1}/{\sqrt{N}}$ for $i = 1,\cdots, N$, $j = 1,\cdots,N_\mathrm{RF}$. The transmitted signal is $\mathbf{x} = \mathbf{F}_\mathrm{RF}\mathbf{F}_\mathrm{BB}\mathbf{s}$.  The goal is to maximize mutual information $\mathcal{I}(\mathbf{F}_\mathrm{RF},\mathbf{F}_\mathrm{BB}) = \log_2 \mathrm{det}( \mathbf{I}_{{N}_\mathrm{S}} + \frac{\kappa}{N_\mathrm{S}\sigma_n^2} \mathbf{H}\mathbf{F}_\mathrm{RF}\mathbf{F}_\mathrm{BB}\mathbf{F}_\mathrm{BB}^\textsf{H}\mathbf{F}_\mathrm{RF}^\textsf{H}\mathbf{H}^\textsf{H}  )$, where $\mathbf{H}\in \mathbb{C}^{N \times N_\mathrm{R}}$ is the wireless channel matrix, $N_\mathrm{R}$ is the number of antennas at the receiver, $\kappa$ is the average received power, and $\sigma_n^2$ is the noise power~\cite{hybridBF_Heath_Ayach2014Jan}. The hybrid beamforming problem is
	\begin{align}
	\label{problemHybridBF1}
	\maximize_{\mathbf{F}_\mathrm{RF},\mathbf{F}_\mathrm{BB}}  \mathcal{I}(\mathbf{F}_\mathrm{RF},\mathbf{F}_\mathrm{BB}) \hspace{10pt}
	&	\subjectto  \| \mathbf{F}_\mathrm{RF}\mathbf{F}_\mathrm{BB} \|_\mathcal{F} = N_\mathrm{S},
	\nonumber \\
	&
	|[\mathbf{F}_\mathrm{RF}]_{i,j}| = {1}/{\sqrt{N}},
	\end{align}
	which is nonconvex because of constant modulus constraint. The product $\mathbf{F}_\mathrm{RF},\mathbf{F}_\mathrm{BB}$ also makes this problem nonlinear. Recast (\ref{problemHybridBF1})  to an equivalent form by minimizing the Euclidean cost between the hybrid beamformer $\mathbf{F}_\mathrm{RF}\mathbf{F}_\mathrm{BB}$ and the unconstrained baseband-only beamformer $\mathbf{F}_\mathrm{C}\in\mathbb{C}^{N\times N_\mathrm{S}}$ as
	\begin{align}
	\label{problemCom1}
	&\minimize_{\mathbf{F}_\mathrm{RF},\mathbf{F}_\mathrm{BB}}  \|\mathbf{F}_\mathrm{RF}\mathbf{F}_\mathrm{BB}  -  \mathbf{F}_\mathrm{C}\|_\mathcal{F}
	\nonumber\\
	&
	\subjectto  \| \mathbf{F}_\mathrm{RF}\mathbf{F}_\mathrm{BB} \|_\mathcal{F} = N_\mathrm{S},
	\nonumber \\
	&
	|[\mathbf{F}_\mathrm{RF}]_{i,j}| = {1}/{\sqrt{N}}, 
	\end{align}
	where $\mathbf{F}_\mathrm{C}$ is obtained from singular value decomposition of the channel matrix $\mathbf{H}$~\cite{heath_Overview_hybridBF}.	In wideband scenario, subcarrier-dependent (SD) digital beamformers are used, and the resulting signal is transformed to the time domain via inverse fast Fourier transform (IFFT) (Fig.~\ref{fig_AnalogDigitalBeamforming}). Then, subcarrier-independent analog beamformers are employed for all subcarriers because the direction of the generated beam does not change significantly with respect to subcarriers in mmWave~\cite{heath_Overview_hybridBF,thz_jrc_Elbir2021Oct}. The hybrid beamforming problem for a wideband system with $M$ subcarriers is 	\begin{align}
	\label{problemCom1Wideband}
	&\minimize_{\mathbf{F}_\mathrm{RF},\mathbf{F}_\mathrm{BB}[m]}  \|\mathbf{F}_\mathrm{RF}\mathbf{F}_\mathrm{BB}[m]  -  \mathbf{F}_\mathrm{C}[m]\|_\mathcal{F} \nonumber\\
	&
	\quad \subjectto  \| \mathbf{F}_\mathrm{RF}\mathbf{F}_\mathrm{BB}[m] \|_\mathcal{F} = MN_\mathrm{S}, \nonumber\\
	& \hspace{20pt} |[\mathbf{F}_\mathrm{RF}]_{i,j}| = {1}/{\sqrt{N}},
	\end{align}
	where $\mathbf{F}_\mathrm{BB}[m]$ is the SD digital beamformer corresponds to the $m$-th subcarrier, $m \in \mathcal{M} = \{1,\cdots, M\}$.

	\begin{figure*}[t]
		\centering
		\subfloat[]{\includegraphics[draft=false,width=.30\textwidth]{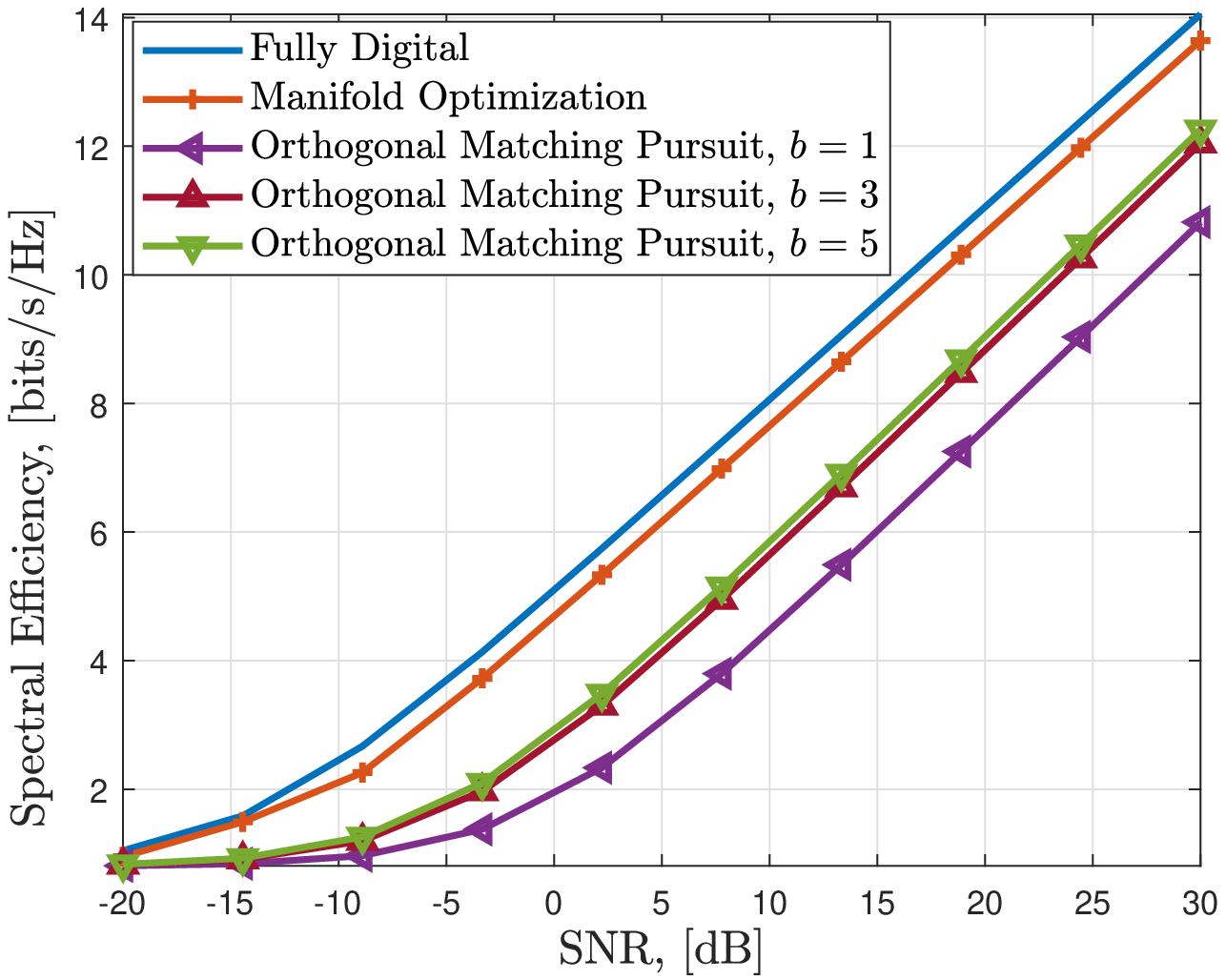} } 
		\subfloat[]{\includegraphics[draft=false,width=.30\textwidth]{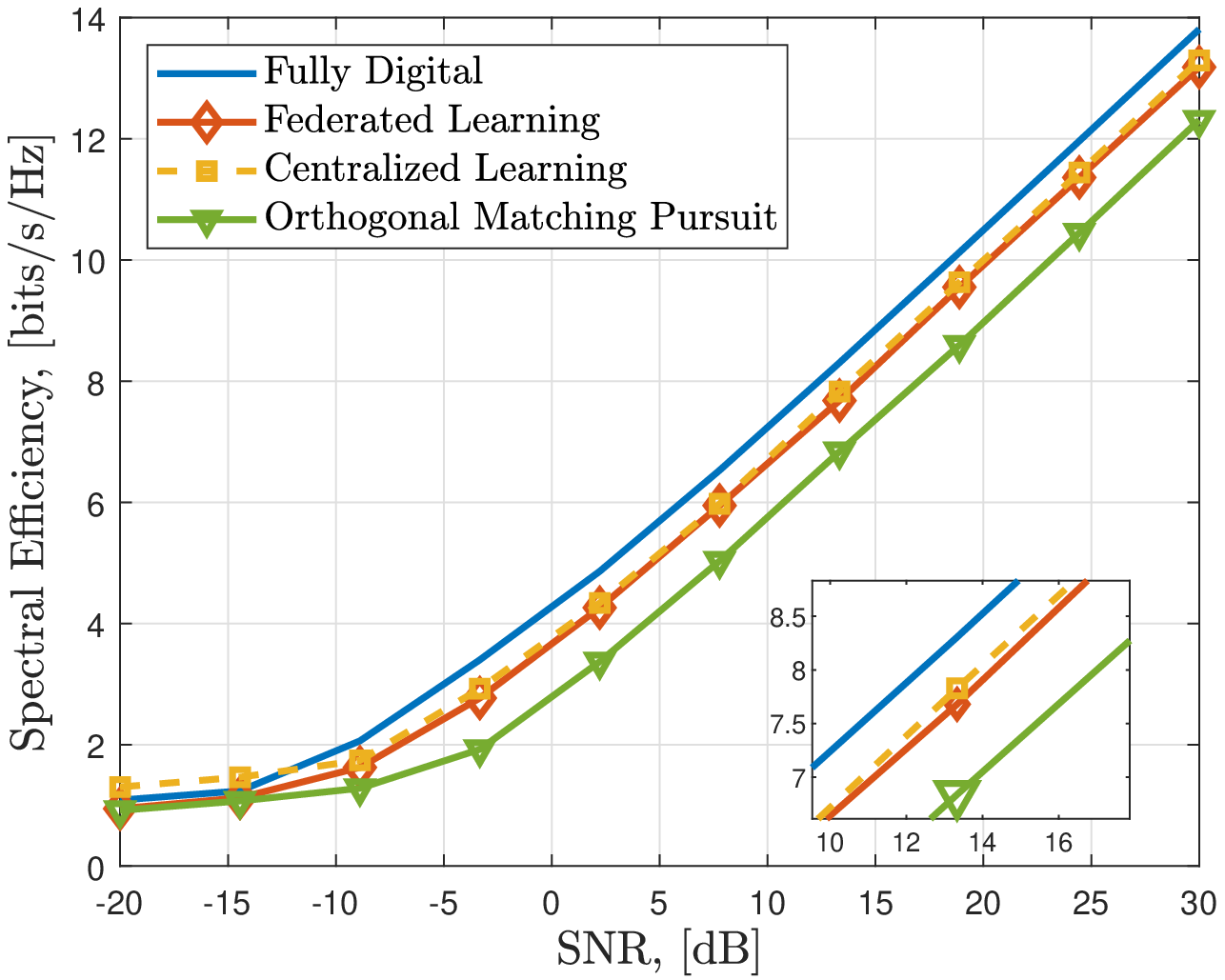} } 
		\subfloat[]{\includegraphics[draft=false,width=.30\textwidth]{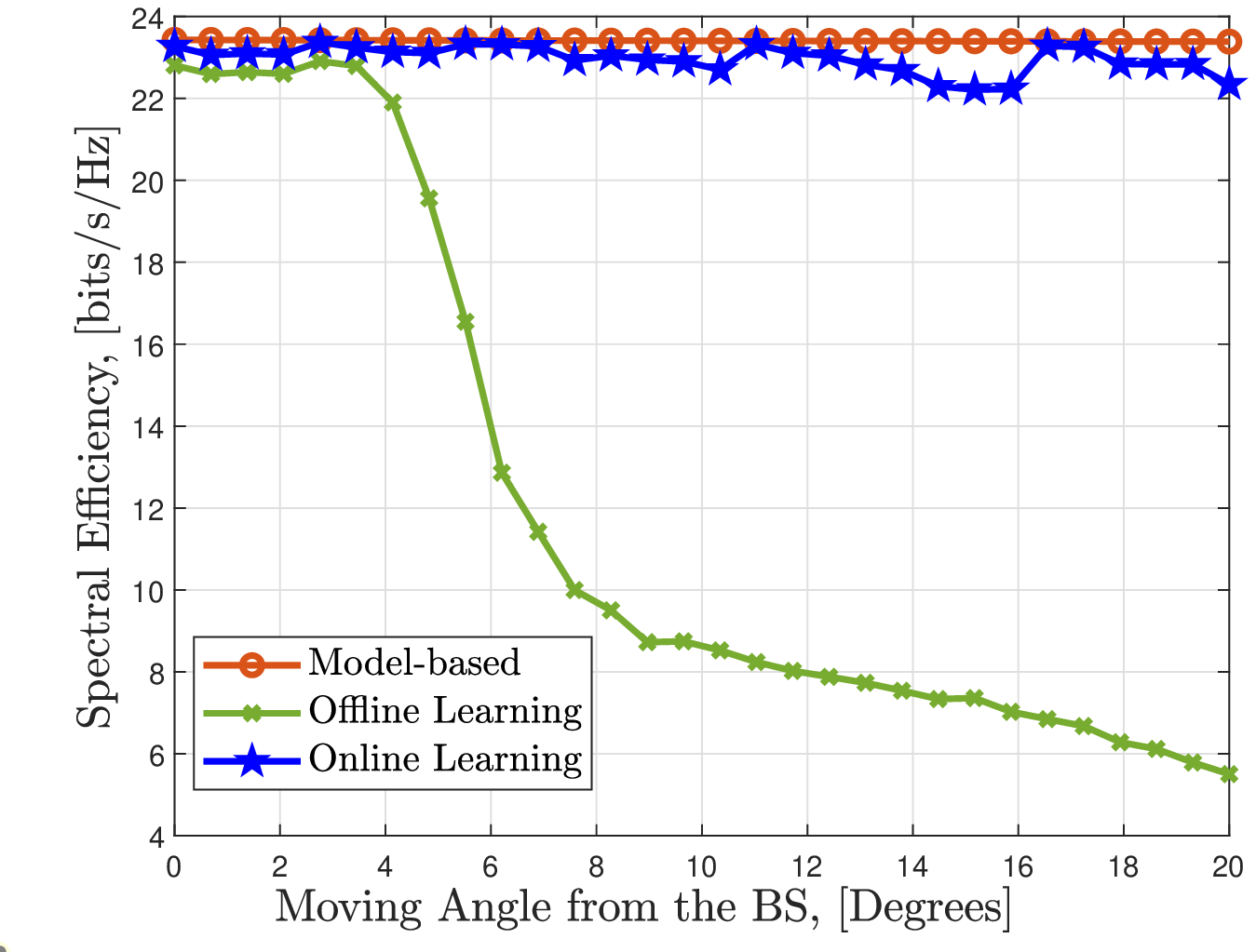} } 
		\caption{\color{black}SE performance of various hybrid beamforming approaches: (a) Low-resolution phase shifters (b) Learning-based and model-based techniques (c) Offline and online learning. Here, the channel is realized with $3$ paths, the number of BS antenna elements $N=100$, the number of users $U=8$, and the number of user antennas $N_\mathrm{R}=16$. 
		}
		\label{fig_LearningVModelBased}
	\end{figure*}
	
	For nonconvex hybrid beamforming formulated in (\ref{problemCom1}), the traditional route is to alternately optimize each ($\mathbf{F}_\mathrm{RF}$ and $\mathbf{F}_\mathrm{BB}$) beamformer iteratively while keeping the other one fixed~\cite{hybridBFAltMin,hybridBF_Heath_Ayach2014Jan,sohrabi_JSTSP}. This has been shown to provide satisfactory SE performance, often close to that of digital-only beamformers, i.e., $\mathbf{F}_\mathrm{C}$~\cite{hybridBFAltMin, hybridBF_Heath_Ayach2014Jan}. During these alternations, while estimation of digital beamformer $\mathbf{F}_\mathrm{BB}$ is straightforward as $\mathbf{F}_\mathrm{BB} = \mathbf{F}_\mathrm{RF}^\dagger \mathbf{F}_\mathrm{C}$, the analog beamformer $\mathbf{F}_\mathrm{RF}$ is difficult to obtain. Often $\mathbf{F}_\mathrm{RF}$ is obtained in terms of the steering vectors via CS-based techniques \textcolor{black}{e.g., orthogonal matching pursuit (OMP). Here, a dictionary of possible steering vectors or atoms is employed and the beamformers are iteratively selected from these atoms based on the similarity between the dictionary and the measurements (i.e., channel data)~\cite{hybridBF_Heath_Ayach2014Jan}}.   \textcolor{black}{In manifold optimization (MO)-based approaches~\cite{hybridBFAltMin}, the search space of $\mathbf{F}_\mathrm{RF}$ is regarded as a \textit{Riemannian submanifold} of $\mathbb{C}^N$ with a complex circle manifold to account for the constant-modulus constraint. Then, the analog and digital beamformers are alternatingly optimized. This method aims to solve the unconstrained optimization problem $\min_{\mathbf{x}} f(\mathbf{x}), \quad \mathbf{x} \in \mathbb{C}^n$,
		where $f(\mathbf{x})$ is the cost function and vector $\mathbf{x}=\operatorname{vec}\left(\mathbf{F}_{\mathrm{RF}}\right)$. To ensure global convergence, the cost function is defined over the Riemannian manifold $\mathcal{M} = \{\mathbf{x}\in \mathbb{C}^{N} | x_n^*x_n = 1, n = 1,\cdots, N \}$. Then, $\mathbf{x}$ is iteratively computed and the solution becomes $\mathbf{x}_{k+1}=\operatorname{Retr}_{\mathbf{x}_k }\left(-\alpha_k \operatorname{grad} f(\mathbf{x}_k)\right)$, where $\operatorname{Retr}$ is the retraction on $\mathcal{M}$ and $\operatorname{grad} f(\mathbf{x}_k)$ denotes the Riemannian gradient~\cite{hybridBFAltMin}. }
	
	\textcolor{black}{The implementation of hybrid analog/digital beamforming imposes another constraint in the system design, i.e., a limited number of phase shifters and analog-to-digital converters (ADCs). Although the power consumption of a phase shifter is typically lower than that of baseband beamformers, their number increases with the number of antennas.  The implementation of hybrid analog/digital beamformers becomes more complex and expensive at higher frequencies (e.g., upper mmWave and THz). As an alternative, lens-based beamformers have been proposed \cite{Abbasi2019Mar}.  Instead of using a phase shifter network, they use lenses to generate a directional beam from the electromagnetic (EM) sources placed at the focal points of the lenses. Thus, lens-based beamformers offer reduced computational complexity when compared with phase-shifter-based architectures. Lens-based beamformers, though, only realize directional beams and not more sophisticated beam patterns as may be useful in a spatial multiplexing or interference cancellation setting. A low-power design in \cite{FazalAsim2022Aug} suggests using Butler matrices, which consist of an $N\times N$ matrix of hybrid couplers and fixed phase shifters.   }

	{\color{black}
		\subsubsection*{Low-resolution ADCs}
		Low-resolution (1-3 bits) ADCs for digital beamformers bring down the overall power consumption and hardware cost. In particular, one-bit ADCs do not require hardware components such as automatic gain control and linear amplifiers. Hence, the corresponding RF chain is implemented cost-efficiently~\cite{alkhateeb2014mimo}. Denote the received signal at the receiver and the corresponding beamformer matrix to be $\mathbf{r}\in \mathbb{C}^{N_\mathrm{R}}$ and  $\mathbf{W}_\mathrm{RF}\in\mathbb{C}^{N_\mathrm{R}\times N_\mathrm{S}}$, respectively. Then, the received signal sampled by low-resolution ADCs is $		\mathbf{r}_q = Q_b(\mathbf{W}_\mathrm{RF}^\textsf{H}\mathbf{r})$,		where $Q_b(\cdot)$ is the quantization operator with $b$ bits resolution. The received signal $\mathbf{r}_q$ is then used to design the receiver via zero-forcing (ZF) or maximum-rate combining (MRC) techniques~\cite{oneBit_CE_Li2017May,alkhateeb2014mimo}.
	}
	
	\subsubsection*{Finite resolution phase shifters}
	In practice, continuous-valued phase angles are expensive to implement and finite resolution phase shifters may be used with low-resolution ADCs. Here, the beamformer weights are selected from the finite set $  \mathcal{W} = \{  1, \omega, \omega^2,\cdots, \omega^{2^b-1} \}$, where $\omega = \frac{1}{\sqrt{N}}e^{\mathrm{j}2\pi/2^b}$ and $b$ is the number of bits. Then, the constant-modulus constraint in (\ref{problemCom1}) is replaced by $[\mathbf{F}_\mathrm{RF}]_{i,j}\in \mathcal{W}$. A feasible solution to hybrid beamforming with finite resolution is to first solve (\ref{problemCom1}) under infinite resolution assumption and then quantize the phase elements of the beamformers~\cite{sohrabi_JSTSP}.
	
	Fig.~\ref{fig_LearningVModelBased}(a) shows the comparison of fully digital beamforming and hybrid beamforming with low resolution phase shifters. The hybrid architecture with MO-based design has performance very close to fully digital beamformers. The OMP with $b=5$-bit phase shifters performs closest to infinite resolution phase shifters. The gap from the fully digital performance is larger for OMP-based techniques compared to MO-based beamforming.

		\begin{strip}
	\begin{tcolorbox}[colback=teal!10,title={}]
		\begin{table}[H]
			\centering
			\vspace{-1em}
			\caption{Learning Models}
			\vspace{-1em}
			\begin{tabular}{p{4.5cm} p{1.7cm} p{8.0cm}}
				\hline
				\textbf{Network model} & \textbf{Data} &  \textbf{Application in beamforming}\\
				\hline 
				\begin{minipage}{0.2\columnwidth}
					\centering
					\includegraphics[width=1.0\linewidth]{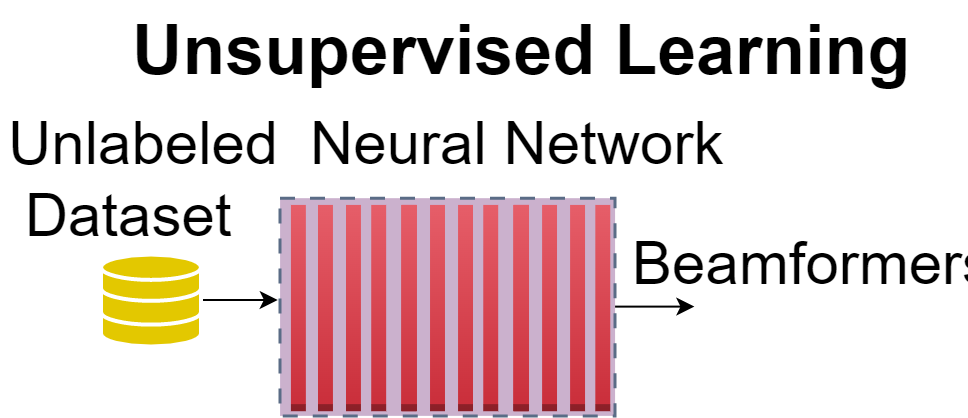}
				\end{minipage} & Unlabeled & \begin{minipage}{0.55\columnwidth}
					\vspace{0.3em} \textit{Fast beamforming:} Minimize a given optimization objective to implicitly obtain the beamformers. UL is useful, especially for mobile transmitters, where labels are not available.
					\vspace{0.3em}
				\end{minipage}\hspace{0.1em}\\
				\hline\begin{minipage}{0.2\columnwidth}
					\centering
					\includegraphics[width=1.0\linewidth]{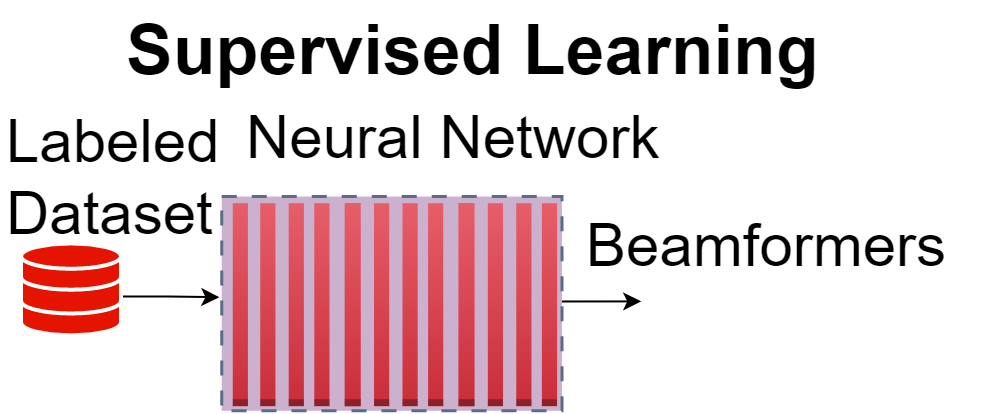}
				\end{minipage} & Labeled & \begin{minipage}{0.55\columnwidth}
					\vspace{0.3em} \textit{Uplink/downlink beamforming:} The network is trained to construct a non-linear relationship between the input and the labeled data (beamformers).  \vspace{0.3em}
				\end{minipage}\hspace{0.1em}\\
				\hline\begin{minipage}{0.2\columnwidth}
					\centering
					\includegraphics[width=1.3\linewidth]{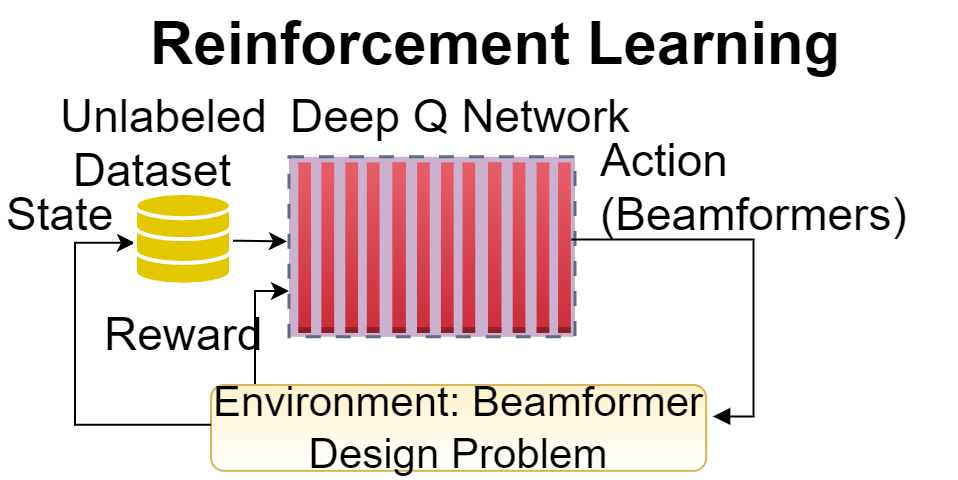}
				\end{minipage} & Unlabeled & \begin{minipage}{0.55\columnwidth}
					\vspace{0.3em} \textit{Uplink/downlink beamforming:} The network learns the beamformers based on a reward/punishment mechanism in accordance with  optimizing the overall system's SE. \vspace{0.3em}
				\end{minipage}\hspace{0.1em}\\
				\hline\begin{minipage}{0.2\columnwidth}
					\centering
					\includegraphics[width=1.3\linewidth]{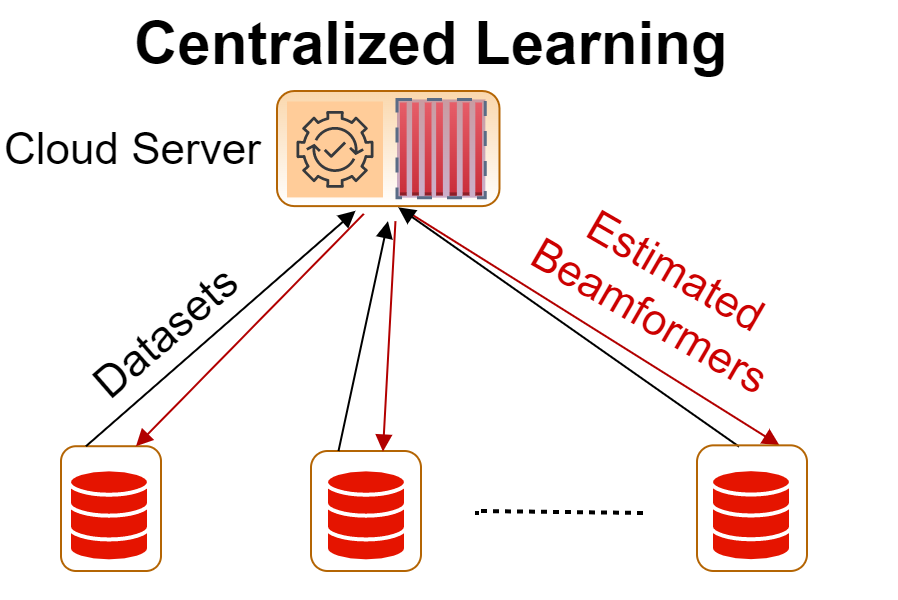}
				\end{minipage} & Labeled/ Unlabeled & \begin{minipage}{0.55\columnwidth}
					\vspace{0.3em} \textcolor{black}{\textit{Uplink multi-user beamforming:} The training datasets are transmitted to a centralized cloud server, wherein the model is trained. Post-training, each user sends the input data (channel) to the server that sends the output (estimated beamformers) to the users.} \vspace{0.3em}
				\end{minipage}\hspace{0.1em}\\
				\hline\begin{minipage}{0.2\columnwidth}
					\centering
					\includegraphics[width=1.3\linewidth]{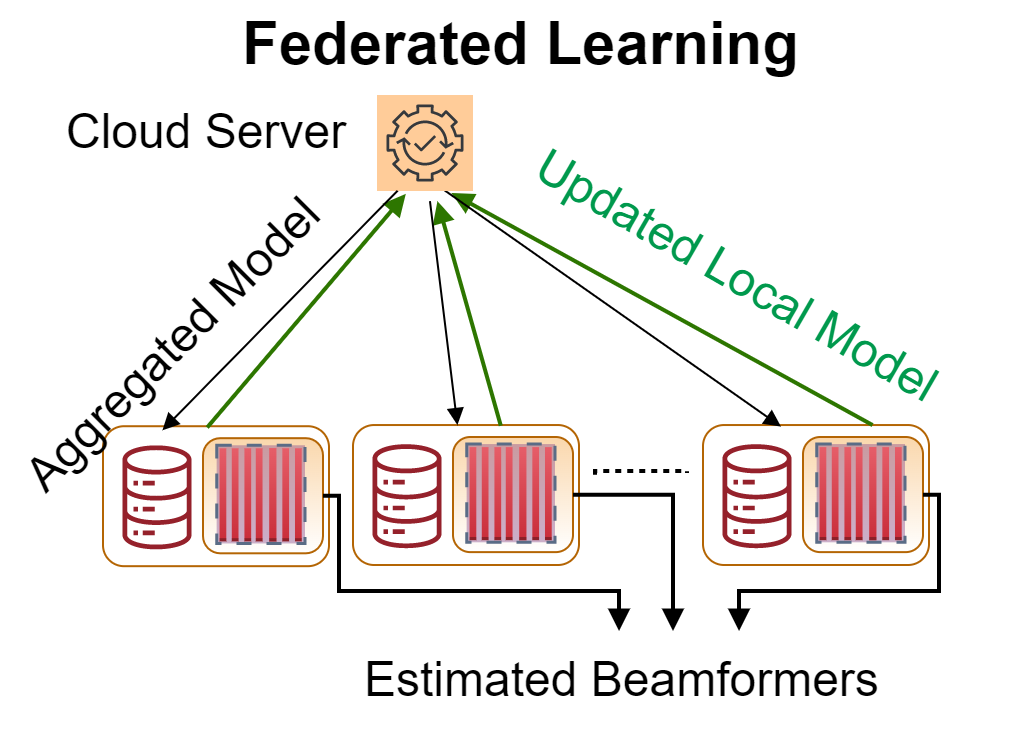}
				\end{minipage} & Labeled/ Unlabeled & \begin{minipage}{0.55\columnwidth}
					\vspace{0.3em} \textit{Downlink multi-user beamforming:} Instead of transmitting the whole dataset to the cloud server, each user processes its own local dataset, computes the corresponding model update, and transmits only the updates to the PS. Then, the server broadcasts the aggregated model updates to the users, which can estimate their own beamformers. \vspace{0.3em}
				\end{minipage}\hspace{0.1em}\\
				\hline\begin{minipage}{0.2\columnwidth}
					\centering
					\includegraphics[width=1.5\linewidth]{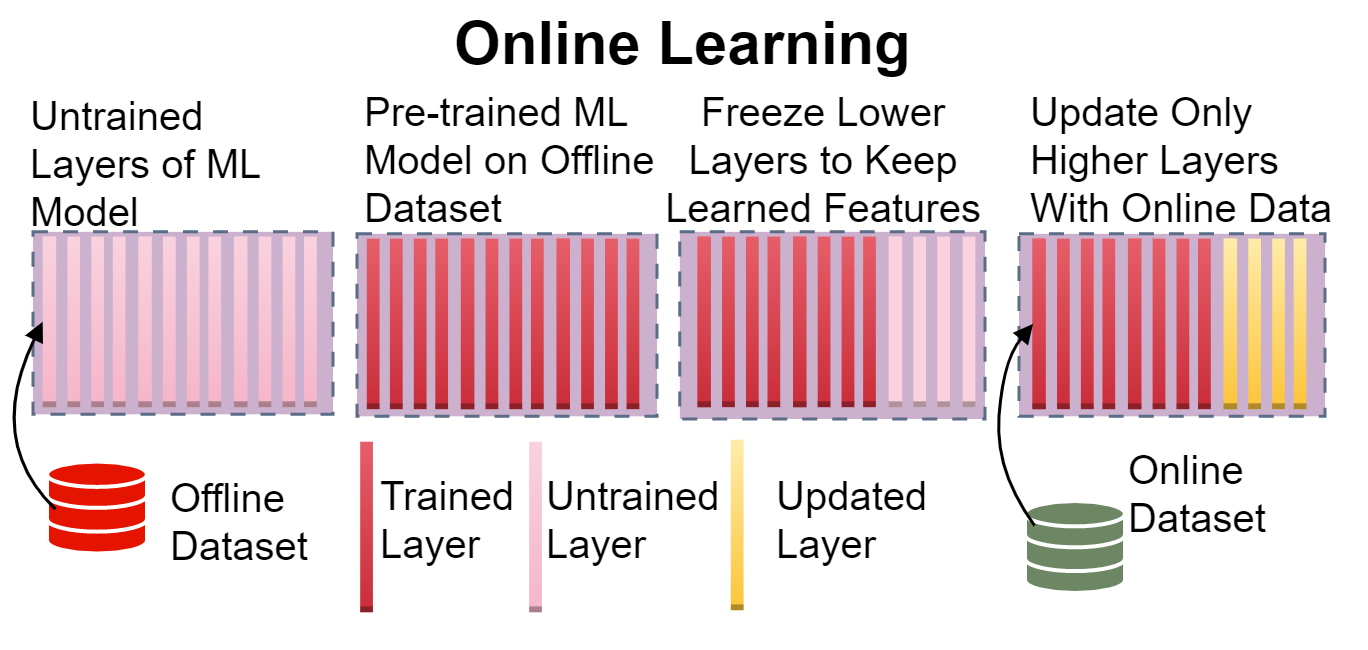}
				\end{minipage} & Labeled & \begin{minipage}{0.55\columnwidth}
					\vspace{0.3em} \textit{Adaptive beamforming:} The learning model is updated when the prediction performance degrades because of deviations in the input compared to the training data.\vspace{0.3em}
				\end{minipage}\hspace{0.1em}\\
				\hline
			\end{tabular}
			\label{table_LearningDiag}
		\end{table}
		\vspace{-2em}
	\end{tcolorbox}
	
		\end{strip} 

	\section{Learning-based beamforming}
	\label{sec:Learning4Beamforming}
	
	Lately, as has been the case with many signal processing problems, beamforming has also not remained untouched by ML techniques. In learning-based hybrid beamforming, the problem is approached from a model-free viewpoint by constructing a non-linear mapping between the input data (e.g., channel matrix, array output) and output (beamformers) of a learning model~\cite{deepL_robustBeamforming_Mohammadzadeh2022Oct,elbir2021DecAFamily,elbirQuantizedCNN2019}. This method has following advantages over model-based techniques: i) The model-free/data-driven structure of a learning-based approach yields a robust performance \textcolor{black}{in terms of SE} against the corruptions \textcolor{black}{(e.g., mismatched number of received paths, imperfectly estimated channel gain and path directions~\cite{elbir2021DecAFamily,elbirQuantizedCNN2019})} in the input. ii) Learning techniques extract feature patterns in the data. Hence, they easily update incoming/future data and adapt in response to environmental changes. \textcolor{black}{The model-based beamformers lack these abilities and may employ statistical predictive algorithms (see Fig.~\ref{fig_LearningVModelBased}c)}; iii) Learning exhibits lower computational complexity in the prediction stage than optimization. Through parallel processing, ML significantly \textcolor{black}{($\sim$10-fold~\cite{elbir2021DecAFamily})} reduces the computational times. On the other hand, a parallel implementation of conventional convex/nonconvex optimization-based beamforming is not straightforward. Beginning from the earlier simpler networks such as multi-layer perception (MLP) to more complex deep learning models like convolutional neural networks (CNNs), ML has come a long way in successfully performing feature extraction for analog and digital beamformers~\cite{elbirDL_COMML}. Table.~\ref{table_LearningDiag} summarizes various learning models, including the well-known unsupervised/supervised learning (UL/SL) and the more recent federated learning (FL).

	\subsection{Unsupervised, supervised, and semi-supervised learning} 
	\label{subsec:ulsl}
	The UL studies the clustering of the unlabeled data into smaller sets by exploiting the hidden features/patterns	derived from the dataset, for which an answer key (label) is not provided beforehand. Hence, the ``distance" between the training data samples is  optimized without prior knowledge of the ``meaning" of each clustered set. \textcolor{black}{In SL, however, the labeled data are used for model training while minimizing the error between the label and the model's response. The cost function of the training is generally the MSE but other functions (e.g., mean error, mean absolute error, cross-entropy, and Kullback-Leibler divergence) may also be used. Note that beamforming may be cast as either a regression (the output is the beamformer weights) or a classification (the output is an index of a vector from a predefined set of possible beamformers) problem.   }   SL is widely used for several applications of beamformer design in radar and communications~\cite{thz_jrc_Elbir2021Oct}.
	
	Define $\mathcal{X}\in \mathbb{R}^{N_\mathrm{in}}$ and $\mathcal{Y}\in \mathbb{R}^{N_\mathrm{out}}$ as the input and label data of a learning model, whose real-valued learnable parameters are stacked into  the vector $\boldsymbol{\Theta}\in \mathbb{R}^Q$. Then, the relationship between the input $\mathcal{X}\in \mathbb{R}^{N_\mathrm{in}}$ and output $\mathcal{Y}\in \mathbb{R}^{N_\mathrm{out}}$ is represented by a nonlinear function $f(\boldsymbol{\Theta},\mathcal{X}): \mathbb{R}^{N_\mathrm{in}}\rightarrow \mathbb{R}^{N_\mathrm{out}} $ such that $	\mathcal{Y} = f(\mathcal{X}|\boldsymbol{\Theta})$.
	The input data are, say, the vectorized elements of the channel matrix $\mathbf{H}$ as $	\mathcal{X} = [\mathrm{vec}\{\Re\{ \mathbf{H}\}^\textsf{T},\mathrm{vec}\{\Im\{ \mathbf{H}\}^\textsf{T}  \}]^\textsf{T},$
	and the labels are beamformers. In the case of the unit-modulus constraint, it suffice to represent the beamformers in terms of only the angle, i.e., $	\mathcal{Y} = \angle \{\mathbf{F}_\mathrm{RF}\}.$ Note that the baseband beamformers are readily computed as $\mathbf{F}_\mathrm{BB} = \mathbf{F}_\mathrm{RF}^\dagger\mathbf{F}_\mathrm{C}$~\cite{hybridBF_Heath_Ayach2014Jan}.
	
	Apart from hybrid beamforming, ML techniques have been applied to other applications such as robust beamformers \cite{deepL_robustBeamforming_Mohammadzadeh2022Oct}. Here, the sample covariance matrix is fed to a CNN whose output is the beamformer weights. The labels are obtained by solving the robust Capon beamformer problem in (\ref{robustCaponBF}). The training dataset was $\mathcal{D} = \{\mathcal{D}_1,\cdots, \mathcal{D}_J\}$, where $\mathcal{D}_i= (\mathcal{X}_i,\mathcal{Y}_i)$ denotes the $i$-th input-output sample for $i = 1,\cdots, J$. The model is trained by minimizing the MSE cost \textcolor{black}{$\frac{1}{J}\sum_{i = 1}^{J}||\mathcal{Y}_i -  f(\mathcal{X}_i|\boldsymbol{\Theta} )  ||_2^2$} over $\boldsymbol{\Theta}$. Post-training, the learned parameters are used for prediction purposes for beamforming. 
	
	The acoustic beamformers in \cite{wager2020fully} are obtained via semi-supervised learning (SSL), where both labeled and unlabeled data are used. When a small set of labeled data are available in addition to a large volume of unlabeled data, using both sets in SSL is more advantageous than SL alone.

	\subsection{Reinforcement learning} In RL, the learning model is initialized from a random state and the algorithms learn to react to the channel conditions on their own \cite{precoderNet_HB_RL}. The model accepts the analog and baseband beamformers of the previous state as input and then updates the model parameters by taking into account the corresponding average rate as a reward. In general, RL has autonomous AI agents that gather their own data and improve based on their trial-and-error interaction with the environment. It shows a lot of promise in basic research. However, so far RL has been harder to use in real-world beamformer applications \textcolor{black}{because its dataset does not include labels. Consequently, RL requires longer training times for learning the features of wireless channels, especially in dynamic, short coherence time scenarios.}

	\subsection{Online learning} 
	
	The OL algorithm involves a learning model whose parameters are updated when there is a significant change in the received input data. For example, consider beamformer design for wireless communications system (Fig.~\ref{fig_LearningVModelBased}(c)), wherein the user is moving away in DoA domain from the BS. Then, the received array data becomes significantly different than the collected offline training data thereby degrading the network performance. Here, hybrid beamforming and channel estimation may be performed jointly because the beamformer weights are directly related to the channel matrix. Moreover, OL is a suitable choice for this problem~\cite{elbir2021DecAFamily}; it updates the model parameters when the normalized MSE (NMSE) of channel estimates is higher than a predetermined threshold. From Fig.~\ref{fig_LearningVModelBased}(c), the learning model requires re-training every $\sim4^\circ$ for a  massive MIMO scenario.
	
	\subsection{Federated learning} 
	
	FL and centralizing learning (CL) are more suited for multi-user scenarios. Using the same neural network structures, CL has a better performance than FL because the former has access to the whole dataset at once whereas the latter employs decentralized training. The FL is ideal for downlink, wherein the trained model is available to the user at the network edge. As an example, consider a downlink scenario, wherein $U$ communications users collaborate to train a model with learnable parameters $\boldsymbol{\Theta} $ with local datasets $\mathcal{D}^{(u)}=(\mathcal{X}^{(u)},\mathcal{Y}^{(u)})$ for $u = 1,\cdots, U$. Here, the output data $\mathcal{Y}^{(u)}$ are the beamformer weights corresponding to the $u$-th user. The FL-based training problem minimizes the averaged local cost 
	$\min_{\boldsymbol{\Theta}} \frac{1}{U}\sum_{u=1}^U \mathcal{L}_u(\boldsymbol{\Theta})$, 
	where $i = 1,\dots, J_u$, and $J_u = |\mathcal{D}^{(u)}|$ denotes the number of samples in $\mathcal{D}^{(u)}$, over $\boldsymbol{\Theta}$. Different than the cost in Section~\ref{subsec:ulsl}, the local cost  here is \textcolor{black}{$\mathcal{L}_u (\boldsymbol{\Theta})  = \frac{1}{J_u} \sum_{i=1}^{J_u} ||f(\mathcal{X}_i^{(u)}| \boldsymbol{\Theta}) - \mathcal{Y}_i^{(u)} ||_2^2$ }
	for the $u$-th user. This is efficiently solved by iteratively applying  
	gradient descent, which updates the model parameter at the $t$-th iteration as $\boldsymbol{\Theta}_{t+1} = \boldsymbol{\Theta}_t - \eta \frac{1}{U} \sum_{u=1}^{U}\boldsymbol{\beta}_u(\boldsymbol{\Theta}_t),$ where $\boldsymbol{\Theta}_t$ is the computed model parameter vector at iteration $t$, $\boldsymbol{\beta}_u(\boldsymbol{\Theta}_t) = \nabla \mathcal{L}_u(\boldsymbol{\Theta}_t)\in\mathbb{R}^Q$ is the gradient vector, and $\eta$ is the learning rate. Fig.~\ref{fig_LearningVModelBased}(b) compares the performance of FL and CL with model-based techniques such as OMP and the fully digital beamformer in terms of SE~\cite{elbir2020FL_HB}. Both CL and FL outperform OMP but the performance gap between CL and FL increases with the non-uniformity of local dataset.

	\begin{figure*}[t]
		\centering
		{\includegraphics[draft=false,width=\textwidth]{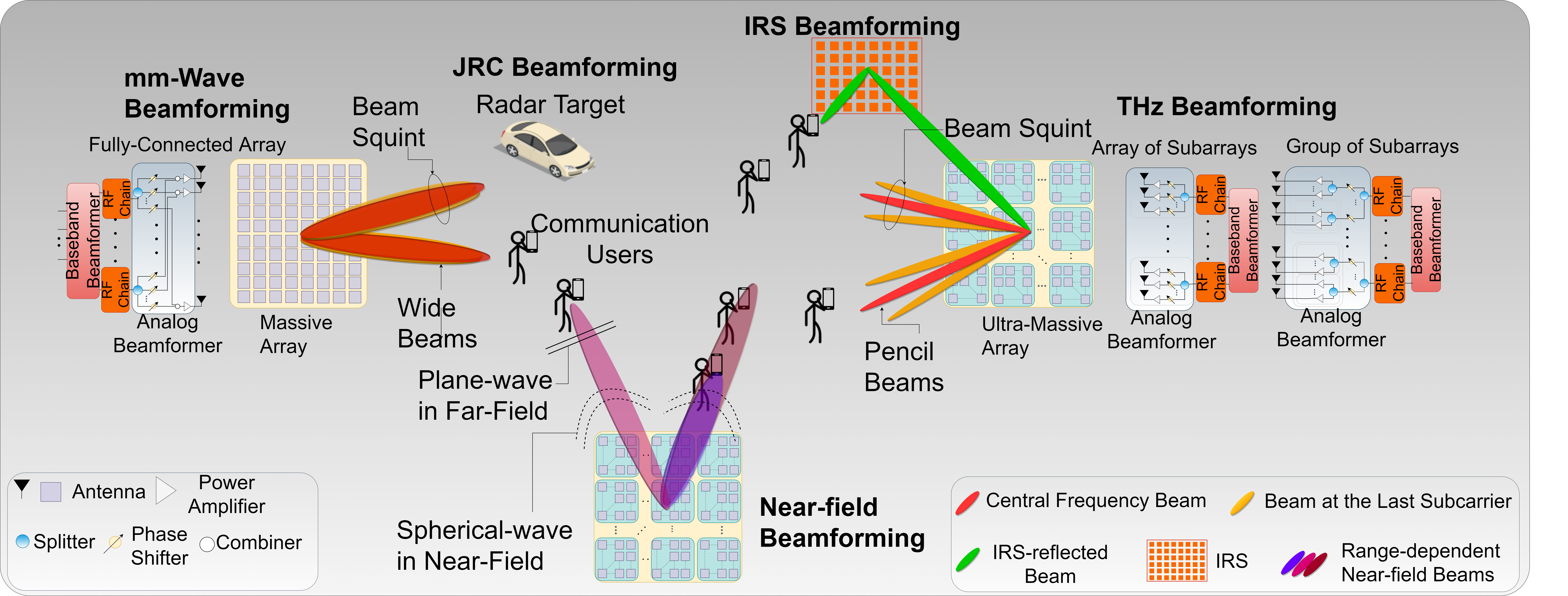} } 
		\caption{A summary of beamforming in emerging applications.  
		}
		\label{fig_BeamformingEmergingApps}
	\end{figure*}

	\section{Emerging Applications}
	\label{sec:EmergingApps}
	Research in beamforming continues to be highly active in light of emerging applications and theoretical advances. For example, the hybrid approach of model-driven network or deep \textit{unfolding} for beamforming \cite{shi2022deep} allows for bounding the complexity of algorithms while also retaining their performance. Convolutional beamformers are gaining salience in acoustics \cite{nakatani2019unified} and ultrasound \cite{heriard2020sparse} as a means to combine multiple, usually non-linear, operations with beamforming. There is also recent interest in beamforming for biomimetic antenna arrays that are based on the direction binaural mechanism of humans or animals \cite{masoumi2013biomimetic,masoumi2013improved}. Synthetic apertures across a wide variety of applications, including quantum Rydberg sensing, present unique beamforming challenges \cite{vouras2022overview}. Holographic beamformers \cite{deng2021reconfigurable} are currently investigated as attractive solutions for multi-beam steering for future wireless applications. In the following, we illustrate a few major applications in the context of radar and communications.

	\subsection{Joint radar-communications}
	For several decades, sensing and communications systems have exclusively operated in different frequency bands to minimize interference with each other at all times. However, this conservative approach for spectrum access is no longer viable because of the demand for wider bandwidth for improved performance of both systems. In the last few years, there has been substantial interest in designing \textcolor{black}{\textit{joint radar and communications}} (JRC) \cite{mishra2019toward} to share the spectrum. From a beamformer design perspective, the problem settings of communications and sensing are combined in JRC. Recall the hybrid beamforming for a communications-only problem as explained in (\ref{problemCom1}). The sensing-only beamformer composed of the steering vectors corresponding to, say, $K$ sensing targets is $\mathbf{F}_\mathrm{R}\in \mathbb{C}^{N_\mathrm{T}\times K}$~\cite{thz_jrc_Elbir2021Oct}. Then, similar to (\ref{problemCom1}), the hybrid beamformer for sensing-only system is obtained by minimizing the Euclidean distance between $\mathbf{F}_\mathrm{RF}\mathbf{F}_\mathrm{BB}$ and $\mathbf{F}_\mathrm{R}\mathbf{P}$ as
	\begin{align}
	\label{problemRadar1}
	&\minimize_{\mathbf{F}_\mathrm{RF},\mathbf{F}_\mathrm{BB},\mathbf{P}}  \|\mathbf{F}_\mathrm{RF}\mathbf{F}_\mathrm{BB}  -  \mathbf{F}_\mathrm{R}\mathbf{P}\|_\mathcal{F} 
		\nonumber\\	&\quad \subjectto    \| \mathbf{F}_\mathrm{RF}\mathbf{F}_\mathrm{BB} \|_\mathcal{F} = N_\mathrm{S}, \;\; 	\nonumber\\	&\hspace{50pt} 
	|[\mathbf{F}_\mathrm{RF}]_{i,j}| = {1}/\sqrt{N}, \;\; \forall i,j,		\;\;\mathbf{PP}^\textsf{H} = \mathbf{I}_{K},
	\end{align}
	where the unitary matrix $\mathbf{P}\in \mathbb{C}^{K\times N_\mathrm{S}}$ is an auxiliary variable to account for different dimensions of $\mathbf{F}_\mathrm{RF}\mathbf{F}_\mathrm{BB}$ and $\mathbf{F}_\mathrm{R}$ without causing any distortion in the radar beampattern. Define $\mathbf{F}_\mathrm{CR}\in \mathbb{C}^{N_\mathrm{T}\times N_\mathrm{S}}$ as the unconstrained JRC beamformer $	\mathbf{F}_\mathrm{CR} = \zeta\mathbf{F}_\mathrm{C}  + (1 - \zeta) \mathbf{F}_\mathrm{R}\mathbf{P}$, where $0\leq \zeta\leq 1$ provides a trade-off between radar and communications performance. Then, the JRC hybrid beamformer is obtained by solving the following optimization problem~\cite{thz_jrc_Elbir2021Oct} 
	\begin{align}
	\label{problem2ISAC}
	&\minimize_{\mathbf{F}_\mathrm{RF},\mathbf{F}_\mathrm{BB},\mathbf{P}} \hspace{3pt} \|\mathbf{F}_\mathrm{RF}\mathbf{F}_\mathrm{BB} -\mathbf{F}_\mathrm{CR} \|_\mathcal{F} 	\nonumber\\	&\quad \subjectto \hspace{0pt} \| \mathbf{F}_\mathrm{RF}\mathbf{F}_\mathrm{BB} \|_\mathcal{F} = N_\mathrm{S},\;\;  
	\nonumber\\	&\hspace{50pt} 
	|[\mathbf{F}_\mathrm{RF}]_{i,j}| = {1}/{\sqrt{N}},\;\; \forall i,j,		\;\; \mathbf{PP}^\textsf{H} = \mathbf{I}_{K}.
	\end{align}
	
		Radar and communications can be combined in other ways, for example leveraging the radar information in a different band to reduce the overheads of configuring the beamforming for communication \cite{AliNGP:PassiveRadar:20}.

	\subsection{THz communications}
	THz-band ($0.1$-$10$ THz) wireless systems have ultra-wide bandwidth and very narrow beamwidth. The signal processing for these systems must address several unique THz challenges, including severe path loss arising from scattering and molecular absorption. In general, THz communications systems employ ultra-massive antenna arrays, which may be variously configured as array-of-subarrays (AoSA) or group-of-subarrays (GoSA)~\cite{thz_jrc_Elbir2021Oct} (Fig.~\ref{fig_BeamformingEmergingApps}), to achieve even higher beamforming gain than mmWave systems.  The wideband beamforming required at THz uses a single analog beamformer for all subcarriers for a hardware-efficient and computationally inexpensive design. However, this leads to beams generated at the lower and higher subcarriers pointing at different directions resulting in \textit{beam-squint} phenomenon~\cite{thz_jrc_Elbir2021Oct}. For comparison's sake, the angular deviation in the beamspace due to beam-squint is approximately $6^\circ$ ($0.4^\circ$) for $0.3$ THz with $30$ GHz ($60$ GHz with $1$ GHz) bandwidth, respectively. One approach to deal with beam-squint is to use time-delayer networks, which is classically known as space-time filtering. Alternatively, one may design a single analog beamformer while passing the effect of beam-squint into the subcarrier digital beamformers. Consider the problem in (\ref{problemCom1Wideband}), where the analog beamformers are subcarrier-independent but the mitigation of  beam-squint implies their SD-ness. Define $\widetilde{\mathbf{F}}_\mathrm{BB}[m]$ as a \textit{beam-squint-aware} digital beamformer. This is obtained via $\widetilde{\mathbf{F}}_\mathrm{BB}[m] = \mathbf{F}_\mathrm{RF}^\dagger \overline{\mathbf{F}}_\mathrm{RF}[m] \mathbf{F}_\mathrm{BB}[m]$, 
	where $\overline{\mathbf{F}}_\mathrm{RF}[m] $ is the SD analog beamformer derived from ${\mathbf{F}}_\mathrm{RF} $ for $m\in \mathcal{M}$ \cite{thz_jrc_Elbir2021Oct}.

	\subsection{Intelligent reflecting surfaces}
	An intelligent reflecting surface (IRS) is composed of a large number of (usually passive)  meta-material elements, which reflect the incoming signal by introducing a predetermined phase shift~\cite{irs_Wu2019Aug}. Thus, IRS-assisted beamforming allows the BS to reach distant/blocked users/targets with low power consumption (Fig.~\ref{fig_BeamformingEmergingApps}). 
	Here, joint optimization of the beamformers at the BS, as well as the phase shifts of IRS elements, is necessary. Consider an IRS-assisted scenario, wherein the IRS is equipped with $N_\mathrm{IRS}$ elements and the BS has $N$ antennas. The transmitted data symbol $s\in \mathbb{C}$ is received at the user as 
	${y}_\mathrm{IRS} = \left( \mathbf{h}_\mathrm{IRS}^\textsf{H}\boldsymbol{\psi} \mathbf{H}_\mathrm{BS} + \mathbf{h}_\mathrm{D}^\textsf{H}  \right)\mathbf{f}s + e$,  
	where $\mathbf{h}_\mathrm{IRS}\in \mathbb{C}^{N_\mathrm{IRS}}$, $\mathbf{h}_\mathrm{D}\in\mathbb{C}^{N}$, and $\mathbf{H}_\mathrm{BS}\in\mathbb{C}^{N_\mathrm{IRS}\times N}$ 
	are the user-IRS, user-BS, and BS-IRS channels, respectively;  the diagonal matrix $\boldsymbol{\psi} = \mathrm{diag}\{[\psi_1,\cdots,\psi_{N_\mathrm{IRS}}]\}\in\mathbb{C}^{N_\mathrm{IRS}\times N_\mathrm{IRS}}$ represents the IRS phase elements; $\mathbf{f}\in \mathbb{C}^{N}$ is the beamformer vector at the BS; and $e\in\mathbb{C}$ is additive noise. The joint active/passive beamformer design becomes
	\begin{align}
	&\maximize_{\boldsymbol{\psi}, \mathbf{f}} \hspace{10pt} | \left( \mathbf{h}_\mathrm{IRS}^\textsf{H}\boldsymbol{\psi} \mathbf{H}_\mathrm{BS} + \mathbf{h}_\mathrm{D}^\textsf{H}  \right)\mathbf{f} |^2 \nonumber\\
	&\subjectto  \| \mathbf{f}\|_2 \leq \bar{p}, 	\hspace{20pt} 0\leq \psi_n \leq 2\pi,
	\end{align}
	where $\bar{p}$ denotes the maximum transmit power and $ n = 1,\cdots, N_\mathrm{IRS}$.

	\subsection{Near-field beamforming}
	Depending on the operating frequency, the wavefront of the transmitted signal appears to have different shapes in accordance with the observation distance. The wavefront is a plane wave in the far-field region. In near-field (Fig.~\ref{fig_BeamformingEmergingApps}), where the transmission range is shorter than the Fraunhofer distance, i.e., $R_\mathrm{NF} = \frac{2 A^2 f_c}{c_0}$, $A$ being the array aperture, the wavefront takes a spherical form. As a result, unlike far-field, the near-field beampattern is range-dependent. For example, the array response vector for ULA is a function of both direction $\theta$ and range $r$ as $\mathbf{a}(\theta,r) = \frac{1}{\sqrt{N}}[e^{-\mathrm{j}\frac{2\pi}{\lambda}  r^{(1)}}, \cdots, e^{-\mathrm{j}\frac{2\pi}{\lambda}  r^{(N)} } ]^\textsf{T}$, 
	where  $r^{(n)} = [r^2 + ((n-1)d)^2 - 2(n-1)dr\sin \theta  ]^{\frac{1}{2}} \approx r - (n-1)d \sin\theta$, ($n=1,\cdots, N$) is a range-dependent parameter corresponding to the receiver and the $n$-th transmit antenna. 
	Hence, the beamformer design needs to account for this spherical model. 

	\section{Summary}
	\label{sec:Summary}
The many beamforming algorithms, their possible variants, and their relative advantages provide a swiss-knife approach to choosing the most appropriate technique for a specific application. We presented an overview of those algorithms that had a considerable impact on signal processing and system design during the last twenty-five years. We focused on radar and communications applications while also mentioning in passing the developments in beamforming for ultrasound, acoustics, synthetic apertures, and optics.

\textcolor{black}{A typical use case of convex beamforming is to allow robustness against various sources of uncertainties such as a small number of snapshots, mismatched SoI direction, and mismatched steering vectors. In nonconvex beamforming, each of the problem settings imposes different constraints on, e.g., PSDness (general-rank beamforming), the probability distribution (chance-constrained robust beamforming), constant-modulus (hybrid beamforming), and received SNR (multicast beamforming).}

\textcolor{black}{Each learning algorithm offers specific advantages of its own. The most common SL (UL and RL) admits labeled (unlabeled) datasets. Furthermore, the inherent reward/punishment mechanism in RL to optimize the learning model for a predefined cost function yields better performance than UL. The FL is particularly helpful for multi-user scenarios whereas CL is preferred if the dataset is small compared to the size of the learning model. When data are updated over time, then OL is beneficial. Note that SL, UL, and RL may also be combined with FL, CL, and OL depending on the problem and data; examples abound such as federated reinforcement learning, online reinforcement learning, online centralized learning, centralized reinforcement learning, and so on.
	}
	

	\section*{Acknowledgments}
	K. V. M. acknowledges support from the U. S. National Academies of Sciences, Engineering, and Medicine via Army Research Laboratory Harry Diamond Distinguished Fellowship.

\balance
	\bibliographystyle{IEEEtran}
	\bibliography{refSPM}

	\begin{IEEEbiographynophoto} 
		{Ahmet M. Elbir}(ahmetmelbir@ieee.org) received the B.S. degree with Honors from Firat University, Turkey, in 2009, and the Ph.D. degree from the Middle East Technical University (METU), Turkey, in 2016, both in electrical engineering. He is currently a research fellow at University of Luxembourg, Luxembourg. His research interests include array signal processing for radar and communications, and deep learning for multi-antenna systems. He serves as an Associate Editor for \textsc{IEEE Access}, and a Lead Guest Editor for \textsc{IEEE Journal of Selected Topics in Signal Processing} and \textsc{IEEE Wireless Communications}. Dr. Elbir is the recipient of 2016 METU Best Ph.D. thesis award for his doctoral studies and the \textsc{IET Radar, Sonar \& Navigation} Best Paper Award in 2022. He is a Senior Member of IEEE.
	\end{IEEEbiographynophoto}
	
	\begin{IEEEbiographynophoto} 
		{Kumar Vijay Mishra} (kvm@ieee.org) received his Ph.D. in electrical and computer engineering and M.S. in mathematics from The University of Iowa while working on the NASA Global Precipitation Measurement Mission ground validation radars. He is a Senior Fellow at the United States DEVCOM Army Research Laboratory and Technical Advisor to startups Hertzwell, Singapore and Aura Intelligent Systems, Boston. He is the recipient of the US National Academies Harry Diamond Distinguished Fellowship and has won many best paper awards. His research interests are radar, remote sensing, signal processing, and electromagnetics. He is a Senior Member of IEEE.
	\end{IEEEbiographynophoto}

	\begin{IEEEbiographynophoto} 
		{Sergiy A. Vorobyov} received the M.Sc. and Ph.D. degrees in systems and control from the National University of Radio Electronics, Kharkiv, Ukraine. He is currently a Professor with the Department of Information and Communications Engineering, Aalto University, Espoo, Finland. He has also held faculty positions with the University of Alberta, Edmonton, AB, Canada and the Joint Research Institute between Heriot-Watt University and Edinburgh University, Edinburgh. His research interests include optimization and multi-liner algebra methods in signal processing and data analysis, statistical and array signal processing, sparse signal processing, estimation, detection and learning theory and methods, and multi-antenna, large-scale, and cognitive systems. Dr. Vorobyov was the recipient of the 2004 IEEE Signal Processing Society Best Paper Award, the 2007 Alberta Ingenuity New Faculty Award, 2011 Carl Zeiss Award, 2012 NSERC Discovery Accelerator Award, and other awards. He is currently serving as the General Co-Chair for EUSIPCO 2023, Helsinki, Finland. He is a Fellow of IEEE.

	\end{IEEEbiographynophoto}
	
	\begin{IEEEbiographynophoto} 
		{Robert W. Heath Jr.} (rwheathjr@ncsu.edu) received he Ph.D. degree from Stanford University in electrical engineering. He is the Lampe Distinguished Professor at  the North Carolina State University. He is also President and CEO of MIMO Wireless Inc. He authored or co-authored several books including Introduction to Wireless Digital Communication (Prentice Hall, 2017) and Foundations of MIMO Communication (Cambridge University Press, 2018). He is the recipient or co-recipient of several awards including the 2019 IEEE Kiyo Tomiyasu Award, the 2020 IEEE Signal Processing Society Donald G. Fink Overview Paper Award, the 2020 North Carolina State University Innovator of the Year Award, the 2021 IEEE Vehicular Technology Society James Evans Avant Garde Award and the 2022 IEEE Vehicular Technology Society Best Vehicular Electronics Paper Award. He was Editor-in-Chief of IEEE Signal Processing Magazine from 2018 - 2020. He is a Fellow of the National Academy of Inventors and a Fellow of the IEEE. 
	\end{IEEEbiographynophoto}

\end{document}